\documentclass[prl,aps,twocolumn,superscriptaddress]{revtex4}
\usepackage{graphicx,color,setspace}

\usepackage{amsmath}
\usepackage[normalem]{ulem}
\begin{document}
\title{The physics of Empty Liquids: from Patchy particles to Water}

\date{\today}

\author{John Russo}
\affiliation{Department of Physics, Sapienza University of Rome, Italy}
\author{Fabio Leoni}
\affiliation{Department of Physics, Sapienza University of Rome, Italy}
\author{Fausto Martelli}
\affiliation{IBM Research Europe, Hartree Centre, Daresbury WA4 4AD, United Kingdom}
\author{Francesco Sciortino}
\affiliation{Department of Physics, Sapienza University of Rome, Italy}

\begin{abstract}
{\bf
Empty liquids represent a wide class of materials whose constituents arrange in a random network through reversible bonds. Many key insights on the physical properties of empty liquids have originated almost independently from the study of colloidal patchy particles on one side, and a large body of theoretical and experimental research on water on the other side.
Patchy particles represent a family of coarse-grained potentials that allows for a precise control of both the geometric and the energetic aspects of bonding, while water has arguably the most complex phase diagram of any pure substance, and a puzzling amorphous phase behavior.
It was only recently that the exchange of ideas from both fields has made it possible to solve long-standing problems and shed new light on the behavior of empty liquids. Here we highlight the connections between patchy particles and water, focusing on the modelling principles that make an empty liquid behave like water, including the factors that control the appearance of thermodynamic and dynamic anomalies, the possibility of liquid-liquid phase transitions, and the crystallization of open crystalline structures.

}
\end{abstract}

\maketitle

In contrast to \emph{simple liquids}, whose properties are determined by hard-core repulsive interactions (and which find in the hard-sphere model~\cite{weeks1971role} or the power-law repulsive pair-potential~\cite{dyre2016simple} accurate reference systems), \emph{empty liquids} represent a class of systems whose properties are based on the formation of a random network through directional bonds.
The definition of empty liquids first appeared in the context of the phase behavior of patchy particle models~\cite{Bianchi_2006,ruzicka2011observation}. Patchy particles represent a family of coarse-grained potentials that, while simplifying the atomistic details of the interactions, allow for a precise control of both the geometric and the energetic aspects of bonding.
Patchy particle models, introduced in the mid-80's as models for associated liquids \cite{Bol_1982,Ghonasgi_1993, Kolafa_1987,A_M_ller_2000,McCabe}, have received a considerable interest in the last decade once it has been realized that they can be quite accurate in modelling the interaction between colloidal particles of new generation with strong directional interactions~\cite{glotzer2007anisotropy, pawar2008patchy,duguet2011design,Bianchi_2006}. They have thus found successful applications in the field of soft matter systems, especially in the colloidal realm~\cite{sciortino2010primitive, Bianchi_2011}.  The interest in patchy models 
has further leavened when it has been realized that these models can  provide a intermediate coarse-grained representation of protein-protein interactions~\cite{liu2007vapor,fusco2013crystallization,skar2019colloid,espinosa2020liquid}, nucleic acid base pairing~\cite{starr2006model}, not to mention smart applications in material design~\cite{romano2020designing,tracey2019programming,rao2020leveraging} and self-assembling into complex target structures~\cite{he2020colloidal}. 

From a  physics  point of view, the renewed interest in patchy colloids has
brought to the attention of the scientific community several unexpected 
phenomena. Without trying to be exhaustive we mention: (i)
the effect of tuning  the valence (the number of possible patch-patch bonds) on the gas-liquid (colloidal poor - colloidal rich) phase separation~\cite{Bianchi_2006}, with the 
remarkable effect that the density of the coexisting liquid approaches zero when the valence approaches two, a concept which is nowadays encoded in the {\it empty-liquid} wording.   (ii) The related opening of 
an intermediate region of densities between the 
coexisting liquid and the dense-glass states and the associated  {\it equilibrium gel}
concept~\cite{sciortino2017equilibrium}, e.g. the formation of  a  highly bonded network forming system without
resorting to the mechanism of arrested phase separation~\cite{lu2008gelation}. (iv) The possibility
to generate equilibrium gel states which have a free energy lower than all possible crystalline structures even at very low temperatures~\cite{smallenburg2013liquids, rovigatti2014gels}. (v) 
The interplay between phase separation, percolation, Fisher lines in the limit of valence approaching two~\cite{stopper2020remnants,tavares2020remnants}. (vi) The possibility to disentangle aggregation and phase separation to clarify their relative role in the thermodynamic description of the system~\cite{janusprl}. (vii) The possibility to design quasicrystal and  open ordered structures through a proper selection of  the number of patches and their angular position~\cite{reinhardt2013computing,van2012formation,tracey2019programming}. In order to guide experiments towards the realization of  colloidal diamond, a tetrahedral crystalline structure with photonic properties~\cite{liu2016diamond,he2020colloidal}, tetrahedral patchy particles models have been extensively studied to uncover the basic assembly principles of these coveted crystals~\cite{zhang2005self,romano2020designing, rao2020leveraging}. 

Many of the new ideas developed in the study of patchy particle models have then found their way back to the study of water~\cite{sciortino2008gel}, which shares a lot of structural similarities with tetrahedral patchy particles, albeit at very different length and energy scales. We like here to build on the original application of patchy colloids as coarse-grained models for associating liquids to focus on water.
Water is a simple molecule, but it is not a simple liquid. It has arguably the most complex phase diagram of any pure substance~\cite{salzmann2019advances}, and a puzzling amorphous phase behaviour~\cite{amann2016colloquium,shephard2017high,martelli_searching2018}, whose connections with the equilibrium phase diagram is a very active research topic~\cite{cerveny2016confined,hestand2018perspective,pettersson2019x,fuentes2019nature,salzmann2019advances,bove2019link,stern2019evidence,martelli_2020,kringle2020reversible,kim2020experimental}. 
Molecules in an almost ideal tetrahedral coordination coexist with molecules exploring  denser and more disordered geometries. As a result of the strong directional hydrogen-bond interactions and the limited number of  bond that a molecule can form, fluctuations in density, energy, entropy assume a temperature dependence that is quite different from the one of simple  (packing controlled)  liquids.  In particular, on supercooling water, response functions increase significantly, a phenomenon which has been connected to  the presence of a metastable liquid-liquid critical point~\cite{poole1992phase}.  

Accurate descriptions of these properties can only be obtained by resorting to computationally expensive models of water~\cite{guillot2002reappraisal,vega2011simulating,gillan2016perspective,morawietz2016van, gartner2020signatures,SerraML2020}, but their complexity often precludes a simple understanding of what makes a liquid behave like water. Luckily, many of the questions that motivate the water community are aimed at understanding water-like behaviour, that is universal to a large family of potentials~\cite{Chandler_2017}. Since the first spectroscopic evidence for the V-shaped geometry of the water molecule was obtained~\cite{mecke1933rotationsschwingungsspektrum}, water-like behaviour has increasingly been associated with that of a random tetrahedral network, held together by strong directional bonds~\cite{sceats1979zeroth} (the hydrogen bonds in the case of water).
The tetrahedral  hydrogen bond network of water  is prone to a particle description based on a limited number (four) of strong directional interactions. The  primitive models of water  (PMW)~\cite{Kolafa_1987}, one of the first example of what we call today a patchy model,  represented water as a hard-sphere decorated by four interaction sites located on a tetrahedral geometry, where the  site-site  interactions  replace the electrostatic charges~\cite{Bol_1982,Ghonasgi_1993,Vega_1998,tu2012different}. 
Despite the short-range nature of the attractive interaction, the molecularly enforced tetrahedral geometry suffices in creating a network that can compare with the geometries revealed by neutron experiments, offering the possibility to shed light on the intimate nature of the anomalous behavior. The Mercedes-Benz models~\cite{Dias_2009,Urbic_2018} and the mW model~\cite{Molinero_2009,Moore_2011} also belong to the class of coarse grained potentials.
The appeal of coarse grained potentials is motivated by their high computational efficiency compared to atomistic models, allowing to considerably extend length and time-scales of simulations making it possible, for example, to study rare events like homogeneous nucleation in direct simulations\cite{lupi2017role,leoni_2019}.
Recently coarse grained models have also been used to interpolate between water-like and simple-liquid behaviours, studying how these properties change with varying one of the potential parameters. For example, the ability to control local tetrahedrality by tuning a single parameter (the tetrahedrality parameter in mW) has opened the door to the study of how water-like behaviour emerges in a continuous way from simple liquid behaviour~\cite{Molinero_2006,russo2018water,ricci2019computational}.

Here we will review some of the key design principles of patchy particles models that have successfully been applied to understand the strange behaviour of water. We first introduce some general properties of patchy particles, highlighting what are the minimal ingredients that are needed to make a fluid behaving like water. In particular we will characterize water as an empty liquid, and study the conditions that give rise to a liquid-liquid transition, a transition between two liquid forms differing in their density. We will then study the connection with water's thermodynamic anomalies, and see how they also emerge naturally from the properties of the underlying tetrahedral network. We will then focus on the crystal phase, and see how the solid-liquid phase transition is affected by the parameters of patchy particles potentials. We will also highlight the conditions that disfavor the crystalline state, and illustrate two important properties that emerge from this suppression, i.e. ultra-stable liquids~\cite{smallenburg2013liquids} and crystal-clear liquid-liquid phase transitions~\cite{smallenburg2014erasing}. We then conclude with an overview on the dynamical properties of patchy particles models, and what they teach us about water's glassy states. Our methodology will make use of simple mean-field methods to describe the general behaviour of these systems, and we will review both numerical and experimental results that first supported these conclusions.

\section{Prologue}

\begin{figure}[t!]
    \centering
    \includegraphics[width=1\columnwidth]{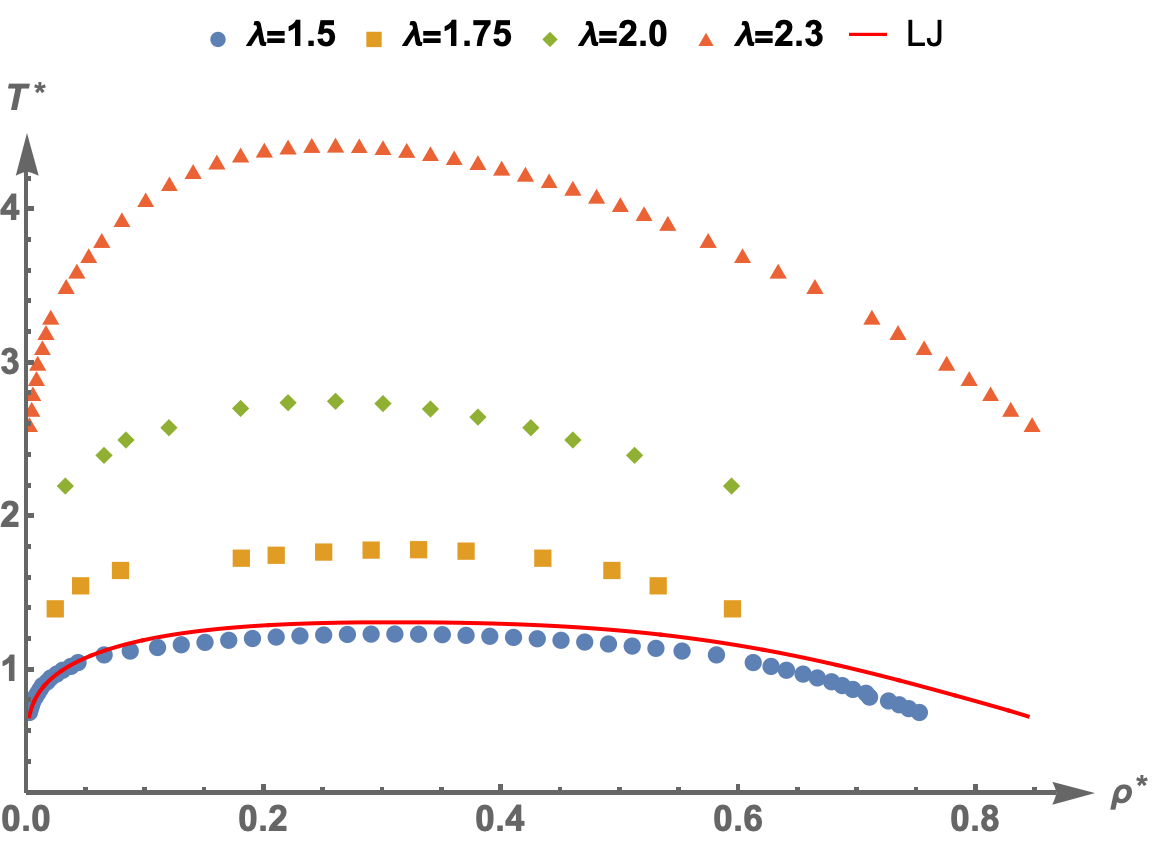}
    \caption{Liquid-gas coexistence lines for the Lennard-Jones (LJ) model (line) and the square-well (SW) model with different values of the well width $\lambda$ (symbols). LJ data is taken from Ref.~\cite{johnson1993lennard}, SW data from Refs.~\cite{del2002vapour,patel2005generalized}, while data close to the critical point is obtained from a cubic fit of the coexistence curve.}
    \label{fig:lj}
\end{figure}

The property that first comes to mind when thinking about the uncommon behaviour of water is the density maximum at $4\,^{\circ}C$ at ambient pressure. The maximum density is around $\rho_\text{m}=1\,g/cm^3$, and it is not as high as the name would suggest. To compare this density with that of other liquids it is convenient to express the number density  $\rho$ in 
non-dimensional units, selecting as unit of length  the nearest neighbour distance  $\sigma$
(the position of the first peak in the radial distribution function) 
The choice  $\sigma=0.28\,nm$,  gives for water $\rho\sigma^3\simeq 0.73$. Typical values for simple liquids are instead of the order of $\rho\sigma^3\simeq 1$. For example, $\rho\sigma^3\simeq  1.05$ for liquid argon at $T=87K$,  $\rho\sigma^3\simeq  1.37$ for benzene, $\rho\sigma^3\simeq  0.95$  for methanol at ambient conditions. In the following we will use non-dimensional units for temperature $T^*$ ($T^*=k_BT/\epsilon$, where $\epsilon$ is the typical interaction energy)
 and scaled number density ($\rho^*=\rho\sigma^3$). %

How hard it is to have liquids with low densities? For particles interacting with spherically symmetric  attractive interactions (in addition to the hard core repulsion), the width of the liquid-gas phase separation at sufficiently low temperatures is quite wide in the $T-\rho$ plane, precluding the possibility of having a low-density stable liquid phase.
To illustrate this point, in Fig.~\ref{fig:lj} we plot the liquid-gas coexistence lines for common spherically symmetric potentials: the Lennard Jones (LJ) potential (continuous line), and the square-well (SW) potential for different values of the width $\lambda$ of the attractive well (symbols). 
The figure shows that for all potentials, and irrespective of the range of the attraction, the liquid branch of the coexistence curve extends to arbitrary high densities, being limited only by the domain of stability of the crystalline state (not shown). The SW model allows us to understand the effect of the range of the attraction, and shows that reducing the range has the effect to drastically suppress the critical temperature, but with a small effects on the critical density. The fact that the liquid branch extends to arbitrary high densities, eventually intersecting the glass transition line~\cite{sastry2000liquid}, is the thermodynamic mechanism for the formation of colloidal gels~\cite{zaccarelli2007colloidal,lu2008gelation,tsurusawa2019direct}.

Thus Fig.~\ref{fig:lj} shows that most isotropic liquids always loose their stability against phase separation at sufficiently low temperatures. 
It is clear that to stabilize liquids at low densities (and low temperatures) a different mechanism is needed. 
This mechanism was first highlighted in the context of patchy particles, where the term \emph{empty liquid} was first introduced~\cite{Bianchi_2006}. 
In the following we will present some basic considerations on the physics of bonding, and then introduce patchy particles and their connection to empty liquids.

\section{Physics of Bonding}

\subsection{The golden rule}

The structure of liquid water is that of a tetrahedral network held together by short-range directional interactions, i.e. hydrogen bonds. From a statistical mechanics point of view, the bonding between two particles can be viewed as a competition between the energy gained from the formation of the bond and the entropy loss due to the reduction in configurational volume that occurs when the two particles are constrained to stay close relative to each other. With $\epsilon$ we will denote the typical bonding energy between two objects. We also define the bonding volume $\mathcal V_\text{b}$ as the volume available for bonding between  two particles. For example, in the case of spherical particles of diameter $\sigma$ interacting with a square well  potential of width $\delta$, the bonding volume is $\mathcal V_\text{b}=4\pi/3 \left[(\sigma+\delta)^3-\sigma^3\right]$, and $\epsilon$ is the depth of the potential well.

The bonding energy in unit of $k_BT$ controls the lifetime of bonds ($\tau$) with an Arrhenius law $\tau\sim \exp(\epsilon/k_BT)$. The bonding volume instead controls the entropy loss per particle after bonding $\Delta s\sim k_B \ln{\mathcal V_\text{b}/(V/N)}=k_B \ln{\rho\mathcal V_\text{b}}$, where $V/N=\rho^{-1}$ is the volume  per-particle. The establishment of an extended network occurs when the competition between energy and entropy is balanced

\begin{equation}\label{eqn:golden_rule}
\rho\mathcal V_\text{b} \exp{\left(\frac{\epsilon}{k_BT}\right)}\sim 1
\end{equation}

For short range attractions, the bonding volume $\mathcal V_\text{b}$ is much smaller than the total volume per particle, which means that an extended network will only form at low values of  $k_BT/\epsilon$.   

\subsection{Patchy particles}

\begin{figure}[t!]
    \centering
    \includegraphics[width=0.8\columnwidth]{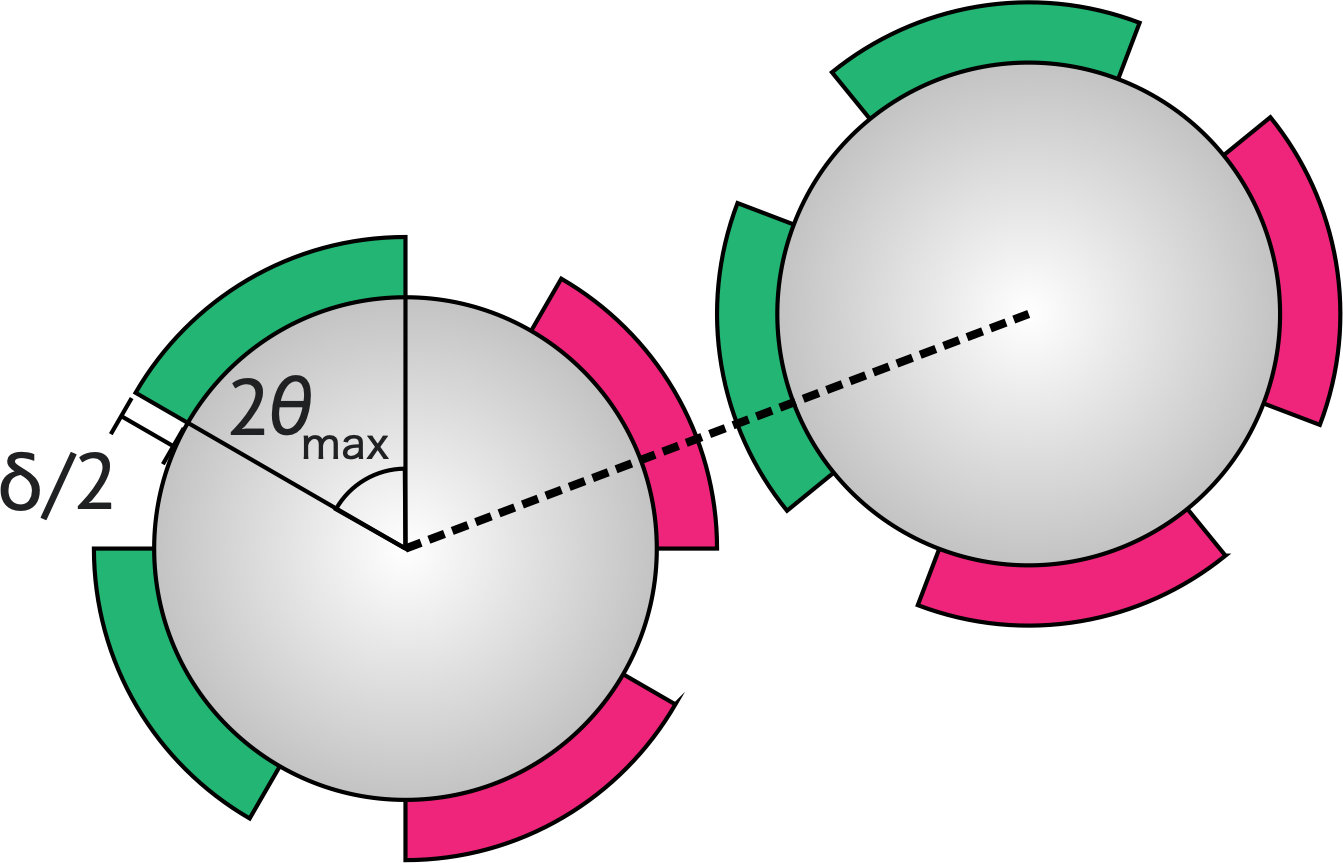}
    \caption{Schematic representation of two patchy particles decorated each with four patches of two different colors. When water is represented as a patchy particle, patches of complementary color (red and green) bind with a  strength $\epsilon$, while same-color patches do not interact. In the Kern-Frenkel model patches interact if the center-to-center distance (dashed line) intersects the area of a patch on each particle. The patch angular width  is controlled by the parameter $\theta_\text{max}$, while the range of the attractive interaction is controlled by $\delta$.}
    \label{fig:patchy}
\end{figure}

The original definition of patchy particles is that of a colloidal particle with attractive spots decorating its surface. Over the years patchy particles have extended beyond the colloidal realm, and have encompassed physical systems where short-range directional interactions play a major role.  Patchy particles models are now being used to offer a coarse-grained description of systems as different as proteins~\cite{liu2007vapor,fusco2013crystallization}, viral capsids~\cite{wilber2007reversible}, hard faceted bodies~\cite{glotzer2007anisotropy,van2014entropically,harper2019entropic}, double-stranded DNA~\cite{largo2007self}, in addition to the early applications in the field of associated liquids, for molecules like water where directional interactions (usually hydrogen bonds) are particularly relevant~\cite{bol1982monte}.

In Figure~\ref{fig:patchy} we plot a schematic representation of two patchy particles interacting through a bond. Each particle is obtained by decorating the surface of a sphere with attractive regions. Different functional forms of the excluded volume interaction and of the directional attraction have been used. For a review on the different models and several computational methods to effectively simulate patchy particles see Ref.~\cite{rovigatti2018simulate}. A popular choice is the Kern-Frenkel potential~\cite{kern2003fluid},
where the repulsive interaction is modelled as a hard-sphere of diameter $\sigma$, while the attractive interaction is described by

\begin{equation}\label{eq:kernfrenkel1}
V_\text{pp}(\mathbf{r}_{ij},\mathbf{\hat{r}}_{\alpha, i},\mathbf{\hat{r}}_{\beta, j})=V_\text{SW}(r_{ij})\,f(\mathbf{r}_{ij},\mathbf{\hat{r}}_{\alpha, i},\mathbf{\hat{r}}_{\beta, j})
\end{equation}

where $\mathbf{r}_{ij}$ is the distance between particle $i$ and $j$, $\mathbf{\hat{r}}_{\alpha, i}$ is the unit vector which points from the center of particle $i$ to the center of patch $\alpha$, and $V_\text{SW}(r_{ij})$ is a square well interaction of range $\sigma+\delta$ and depth $\epsilon$. The orientation-dependent modulation term is described by

\begin{equation}\label{eq:kernfrenkel2}
f(\mathbf{r}_{ij}, \mathbf{\hat{r}}_{\alpha, i}, \mathbf{\hat{r}}_{\beta, j}) = \left\{ 
\begin{array}{rl}  
1 & \mathrm{if} \; \begin{array}{rl} 
\mathbf{\hat{r}}_{ij} \cdot \mathbf{\hat{r}}_{\alpha, i} > \cos{\theta_\mathrm{max}}\\
\mathbf{\hat{r}}_{ji} \cdot \mathbf{\hat{r}}_{\beta, j} > \cos{\theta_\mathrm{max}}
\end{array}\\
0 & \mathrm{otherwise.}
\end{array} \right.
\end{equation}

Thus, this modulation term is different from zero only when the center-to-center distance between two patchy particles intersects the volume of a patch on both particles, as schematically shown in Fig.~\ref{fig:patchy}.

We already introduced the concept of bonding volume $\mathcal{V}_\text{b}$ as one of the key parameters that controls the aggregation of particles. 
In the one-patch Kern-Frenkel model the patch bonding volume has a simple expression

\begin{equation}\label{eqn:bondingvolume}
\mathcal{V}_\text{b}=\frac{4\pi}{3}((\sigma+\delta)^3-\sigma^3)\left(\frac{1-\cos{\theta_\mathrm{max}}}{2}\right)^2
\end{equation}

with parameters describing the width of the angular interaction ($\cos{\theta_\mathrm{max}}$) and the range of the attraction ($\delta$). We will show that (in the single bond per patch condition)  $\mathcal{V}_\text{b}$ is the only parameter controlling the thermodynamics of amorphous phases at the mean-field level. This means that two patchy particles having different geometries of the patch interactions but the same $\mathcal{V}_\text{b}$ will have the same free energy.

\subsection{Mean-field theory of Association}

Associating fluids are those whose molecules can aggregate to form long-lived structures thanks to intramolecular forces that are stronger than typical dispersion interactions but weaker than covalent bonds.
We review here the basic steps to derive a mean-field theory of associating fluids~\cite{hill1986introduction,chapman1988phase,sciortino2019entropy}, 
for identical particles with $M$ functional sites, putting emphasis on the approximations that are involved in its derivation.

The first assumption is that the aggregating units associate into clusters that do not interact with each other and are in thermodynamic equilibrium. This approximation is called the \emph{ideal gas of cluster approximation}
and allows us to write the pressure of the system as that of an ideal gas in which the number of atoms is substituted by the number of clusters
\begin{equation}
\beta P V= N_c
\end{equation}
where $N_c$ is the total number of clusters. Since clusters of any size are in thermodynamic equilibrium with each other, they can be thought of different species  whose composing monomers have the same chemical potential. 
The chemical potential of a cluster of size $n$ is thus $n\mu$, where $\mu$ is the chemical potential of the isolated (not-bonded) monomer, whose value is (assuming for
simplicity that the de Broglie thermal wavelength is one)

\begin{equation}
\beta\mu=\log\rho_1
\end{equation}
with  $\rho_1$ being the number density of isolated monomers.

The second approximation is that every bond in the system reduces the number of clusters by one. In other words a new bond cannot form inside a cluster, but only between two different clusters. This condition is met with the following two requirements: i) two particles cannot be doubly bonded (the one-bond-per-patch condition); ii) bonds within a cluster do not form loops. This is equivalent to an aggregation process on the Bethe lattice, which is a loop-less infinite-dimensional lattice.

Both the number of clusters $N_c$ and the density of monomers can be expressed in terms of the probability that a bond is formed $p_b$. Let's consider a system of $N$ particles all having the same number of patches $M$ (valence). The maximum number of possible bonds is given by $N_{max}=MN/2$, and 
thus $p_b=N_b/N_{max}$ where $N_b$ is the total number of bonds in the system. Within the mean-field approximation, the number of clusters is the total number of particles minus the number of bonds

\begin{equation}
N_c=N-N_b=  N\left(1-M \frac{p_b}{2}\right)
\end{equation}

The number density of monomers (clusters of size one) is proportional to the probability that all $M$ sites of the particle are unbounded

\begin{equation}
\rho_1=\rho (1-p_b)^M
\end{equation}

Putting together the equations above we arrive at an expression for the free energy per particle $f=F/N$ in terms of the bonding probability $p_b$

\begin{eqnarray}
\beta f &=& \beta\mu-\frac{\beta P}{\rho} \nonumber \\
&=&\beta\mu - \frac{N_c}{N} \nonumber \\
&=& \log\rho(1-p_b)^M -\left(1-\frac{M\,p_b}{2}\right)
\end{eqnarray}

This free energy can be divided in ideal and bonding contributions

\begin{eqnarray}
\beta f_\text{id}&=&\log\rho-1 \label{eqn:fid}\\
\beta f_\text{bond} &=& \log(1-p_b)^M +\frac{M\,p_b}{2} \label{eqn:fbond}
\end{eqnarray}

The bonding probability is obtained through a mass-balance equation, i.e. the ratio between the probability of a formed bond, $p_b$, and an open bond, $(1-p_b)^2$, is expressed as an activation function

\begin{equation}
\frac{p_b}{(1-p_b)^2} = \exp\left(-\beta (\Delta U_b-T\Delta S_b)\right)
\end{equation}

Identifying~\cite{sciortino2007self}  $\Delta U_b=-\epsilon$ and $\Delta S_b=k_B\log\left(\rho \mathcal{V}_\text{b}  M)\right)$ we arrive thus at the following equation for the bonding probability

\begin{equation}\label{eqn:pbond}
\frac{p_b}{(1-p_b)^2} = M \mathcal{V}_\text{b}  \rho \exp\left(\beta\epsilon\right)
\end{equation}
where  $ \mathcal{V}_\text{b}$  is  the expression in Eq.~\ref{eqn:bondingvolume}, and $M \mathcal{V}_\text{b}$ is the bonding volume corresponding to $M$ patches.

Equations~\ref{eqn:fid},~\ref{eqn:fbond},~\ref{eqn:pbond} are the basic equations for a mean-field theory of bonding.
While this treatment is of limited accuracy due to the assumptions involved, it provides the correct low-density limit to the free energy.

We now briefly describe some improvements that, while retaining the same conditions we introduced for the mean-field model, considerably improve on the accuracy of the results.

\subsection{Wertheim perturbation theory}
Wertheim~\cite{wertheim1984fluids} introduced a thermodynamic theory in terms of a cluster expansion of monomers and bonded species, which are related to each other by a mass-balance equation. Since it's a perturbation theory it requires the structure of the reference fluid to be known. 
Here we present  the equations for the case of patchy particles where the repulsive interaction is represented by a hard-core interaction, and each particle has $M$ patches. In this case the reference fluid is the hard-sphere model, whose configurational free energy is accurately represented, in the stable fluid region, by the Carnahan-Starling formula

\begin{equation}\label{eqn:carnahan}
\beta f_\text{HS}=\frac{4\phi-3\phi^2}{(1-\phi)^2}
\end{equation}

where $\phi=\rho\sigma^3\pi/6$ is the packing fraction.
The bonding free energy is the same as in the mean-field treatment (Eq.~\ref{eqn:fbond}), but the bonding probability in Eq.~\ref{eqn:pbond} is replaced by 

\begin{equation}\label{eqn:pbond_wertheim}
\frac{p_b}{(1-p_b)^2} = M \rho 4\pi\int_{V_b} g_\text{ref}(r) \left<\exp{(-\beta V(r))-1}\right>r^2\,dr
\end{equation} 

where $g_\text{ref}(r)$ is the radial distribution function of the reference fluid (in this case hard spheres), 
$V(r)$ is the patch-patch interaction potential, and the integration domain is the bonding volume and  $\left< ^{.....} \right >$ indicates a spherical average.

It is easily shown that Wertheim's expression (Eq.~\ref{eqn:pbond_wertheim}) in the limits of low density ($g_\text{ref}(r)\simeq 1$) and low temperature ($\exp{(\beta\epsilon)}\gg 1$) reduces to the mean-field expression in Eq.~\ref{eqn:pbond}.

Wertheim's perturbation theory was originally developed to model associated liquids.
Several modeling choices are possible for water. The simplest one is to have particles with four identical patches so that each particle can have up to four bonds: this choice corresponds to putting $M=4$ in Eq.~\ref{eqn:fbond} and Eq.~\ref{eqn:pbond_wertheim}. Another possibility is to model the water molecule as a particle having four bonds of two different colors (see Fig.~\ref{fig:patchy}) that represent the hydrogens (acceptors) and oxygen's lonely pairs (donors)~\cite{bol1982monte,Kolafa_1987}. Only patches of complementary color can interact, while same-color interactions are forbidden. Due to the symmetry of the bond, the bonding probability of the donor and the acceptor are the same $p_b^\text{donor}=p_b^\text{acceptor}$ and the two color model has the same free energy expression as a function of $p_b$ as the one color model with four patches, Eq.~\ref{eqn:fbond} with $M=4$. The difference between the two models is in the mass-balance Eq.~\ref{eqn:pbond_wertheim}, where the two color model uses $M=2$. The one color model has had a wider adoption and so it will be our model of choice.

We note for completeness that recently there have been attempt to extend the Wertheim theory to relax the assumption of single bond per patch conditions and multiple inter-particle bondings~\cite{sear1994thermodynamic,tavares2012quantitative,marshall2013density,marshall2013thermodynamic,howard2020wertheim}.

\subsection{Empty Liquids}\label{sec:emptyliquids}

\begin{figure}[t!]
    \centering
    \includegraphics[width=1\columnwidth]{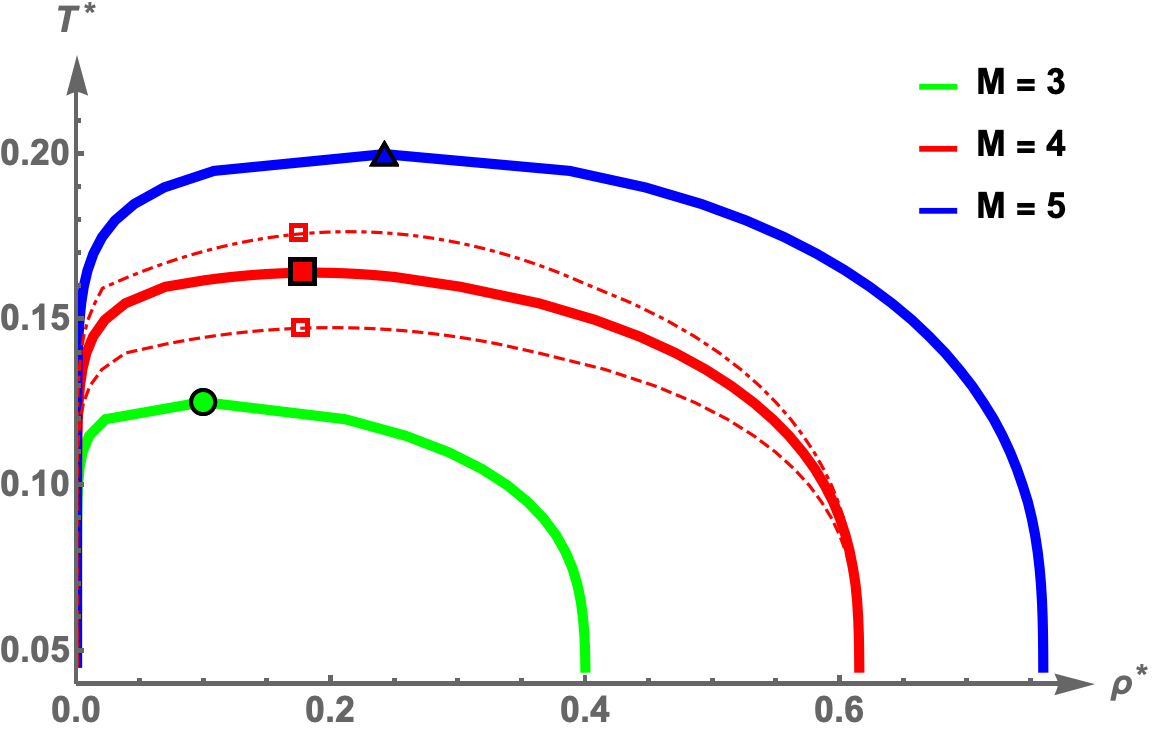}
    \caption{Phase diagram from Wertheim theory for different values of the number of patches $M$. The continuous lines are for $\mathcal{V}_\text{b} =0.005$ and $\epsilon=1$. The dashed and dot-dashed lines show the liquid-gas binodal lines for $M=4$ and $\mathcal{V}_\text{b} =0.0025$ and $\mathcal{V}_\text{b} =0.0075$ respectively.}
    \label{fig:wertheim}
\end{figure}

Fig.~\ref{fig:wertheim} shows the liquid-gas binodal lines predicted from Wertheim theory (Eq.~\ref{eqn:fbond},~\ref{eqn:carnahan},~\ref{eqn:pbond_wertheim}) for different values of the number of patches $M$ and for the bonding volumes 
(continuous lines). Critical points are depicted as points. To ease calculations, we have approximated the integral in Eq.~\ref{eqn:pbond_wertheim} by the value of the pair correlation at contact (that for hard spheres is proportional to the pressure of the reference  system)

\begin{eqnarray}\label{eqn:wertheimbonding}
\int_{\mathcal{V}_\text{b} } g_\text{ref}(r)&& \left<\exp{(-\beta V(r))-1}\right>r^2\,dr \approx  \nonumber \\
&& \mathcal{V}_\text{b} \left(\frac{1-0.5\phi}{(1-\phi)^3}\right) \left(e^{-\beta \epsilon}-1\right)
\end{eqnarray}

The first noticeable effect of reducing the valence is to shift the coexistence curve to lower temperatures and to lower densities. Moreover, differently from the isotropic potentials shown in Fig.~\ref{fig:lj}, the liquid branch falls steeply with density, opening up a region in the phase diagram where liquids are free from phase separation to arbitrary low temperatures. The case $M=4$ shows that tetrahedrally coordinated liquids like water do not suffer from phase separation for densities larger than approximately $\rho^*\sim 0.6$.

In Fig.~\ref{fig:wertheim} we also plot coexistence curves for the $M=4$ and for $50\%$ lower and higher bonding volumes (dash and dash-dot line respectively). We see that the bonding volume has an effect on the critical temperature, but it does not change significantly the critical density.

The possibility of reaching liquids with vanishingly small densities by decreasing the average valence of the system has spurred a lot of research on so-called \emph{Empty Liquids}~\cite{Bianchi_2006,Bianchi_2019}. By mixing particles with different valence one can continuously change the valence also to non-integer numbers. For liquid-gas phase coexistence the limiting case is for $M=2$~\cite{Bianchi_2006,de2011phase,teixeira2017phase}. When the valence is two, particles can only aggregate into chains, for which no phase transition to a spanning network exists. The critical density and temperature go continuously to zero as $M\rightarrow 2$, but surprisingly some thermodynamic loci usually associated with criticality, like the compressibility maxima line, persist in that limit~\cite{stopper2020remnants,tavares2020remnants}.

\section{Liquid-Liquid phase transitions}

\begin{figure}[t!]
    \centering
    \includegraphics[width=1\columnwidth]{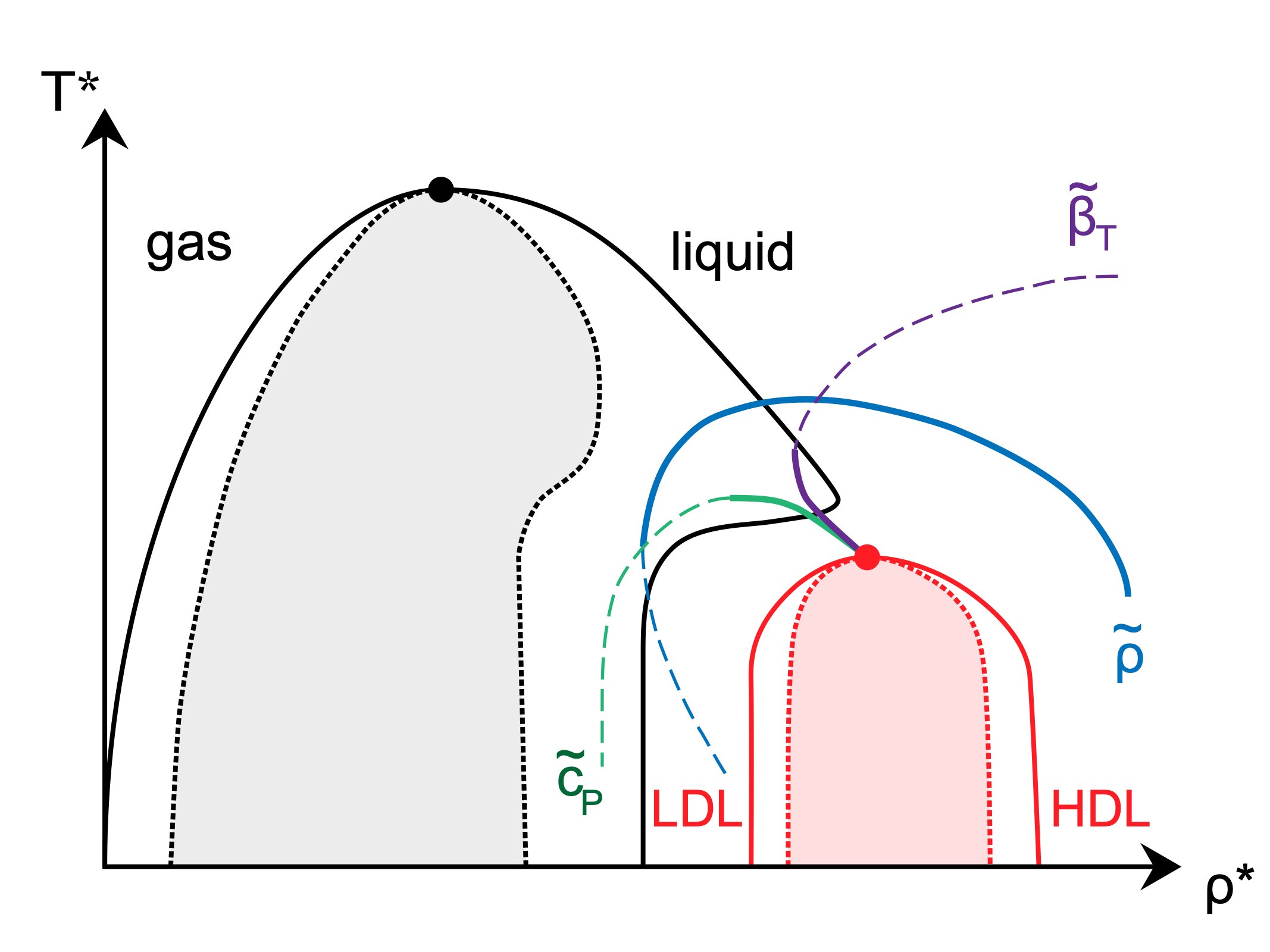}
    \caption{Schematic phase diagram of water in the Liquid-Liquid critical point scenario. In black and red we draw the ordinary gas-liquid and the liquid-liquid phase separations respectively: critical point (full symbol), bimodals (full lines), spinodals (dotted lines). $\tilde{\rho}$, $\tilde{\beta}_T$, $\tilde{c}_P$ are liquid anomaly lines, where the density $\rho$, isothermal compressibility $\beta_T$ 
    and specific heat $c_P$ are extrema. Liquid anomalies are depicted as full lines for maxima of the corresponding response functions and dashed lines when they represent minima.}
    \label{fig:llcp}
\end{figure}

In the previous section we have seen that bonding can give rise to a gas-liquid phase transition~\cite{romano2007gas}. Some of the first successful applications of Wertheim theory were indeed in the modeling of the liquid-gas coexistence curve of water. For a comprehensive review of these efforts see for example Ref.~\cite{clark2006developing}.

In the following we will focus on the supercooled behaviour of water, for which several possible scenarios have been proposed~\cite{speedy1982stability,debenedetti1991spinodal,poole1992phase,gallo2016water}. These scenarios all revolve around the possible existence of a liquid-liquid critical point, below which the liquid phase separates  in two liquids, LDL and HDL, the low and high density liquids respectively. We offer a schematic representation of the proposed phase diagram of water, 
consistent with recent numerical studies~\cite{poole1992phase, poole2005density, palmer2018advances, debenedetti2020second}, 
 in Fig.~\ref{fig:llcp}. The ordinary gas-liquid phase separation is presented in black, and we also plot the critical point (full point), binodals (full lines) and spinodals (dotted lines). At low temperatures, we see the appearance of a second transition (in red), with a liquid-liquid critical point (LLCP, full point) that connects the LDL and HDL binodals (full lines) and spinodals (dashed lines).

The location of the LLCP depends  sensibly on the parameters of the model, and can be moved quite arbitrarily by tuning these parameters. What distinguishes the different scenarios for the supercooled water behaviour is the relative location of the LLCP with respect to the liquid-gas phase transition lines. For example, moving the LLCP inside the gas-liquid spinodal region (black shaded region) gives origin to the \emph{critical point free scenario} \cite{poole1994,angell2008insights}. Moving the critical temperature of the LLCP to zero instead gives the \emph{singularity free scenario}~\cite{sastry1996singularity}. All these scenarios are thus part of the same thermodynamic landscape, and several models that can interpolate continuously between them have been proposed~\cite{russo2018water,rovigatti2017communication,chitnelawong2019stability}.

In the following we will look for conditions that promote liquid-liquid phase separation. Due to their equivalence, we will not focus on the different scenarios, and instead look for models that in an appropriate range of temperatures give rise to two separate regions of instability (grey and pink shaded regions in Fig.~\ref{fig:llcp}).

\subsection{Phase separation in the Van der Waals fluid}

\begin{figure}[t!]
    \centering
    \includegraphics[width=1\columnwidth]{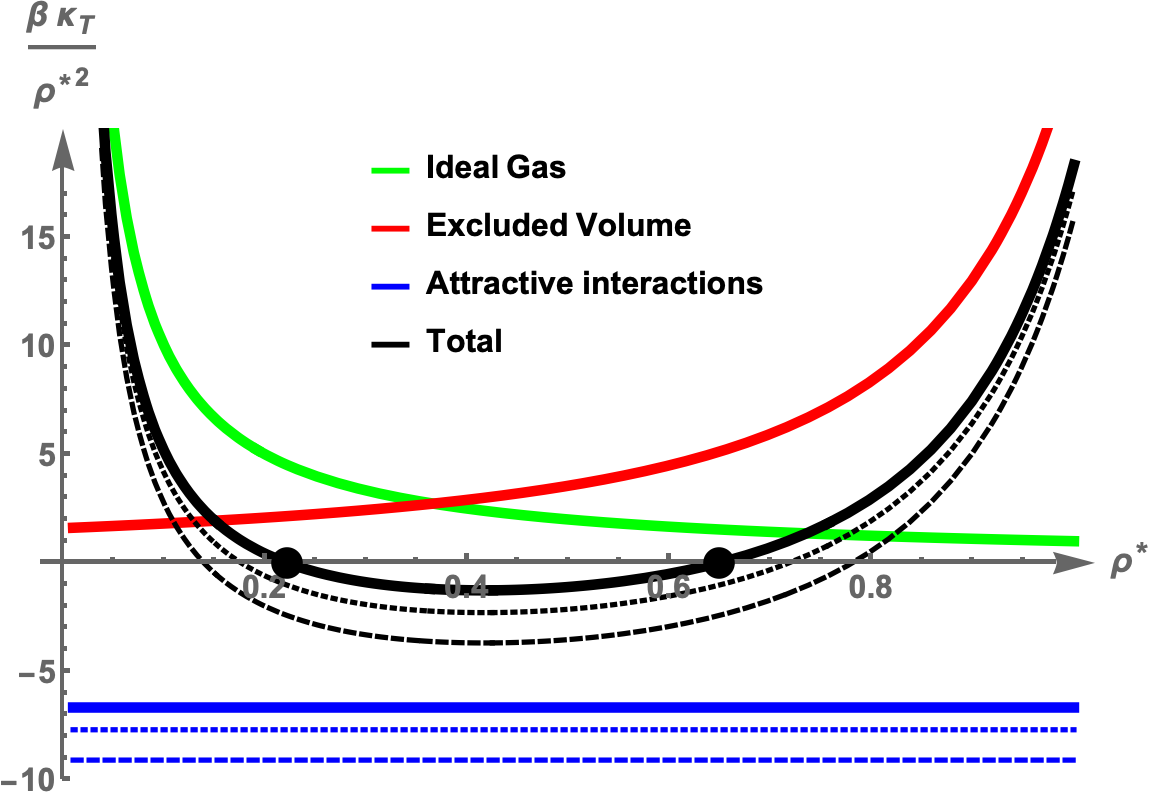}
    \caption{Isothermal bulk modulus, $\beta\kappa_T/\rho^2$, for the Van der Waals fluid with $a=0.5$ and $b=0.8$ for which $T^{*}_c=0.185$. Continuous lines represent the different contributions to the bulk modulus at $T/T_c=0.81$. Full points mark the location of the spinodal points. Dotted and dashed lines present the attractive and total contributions for $T/T_c=0.70$ and $T/T_c=0.59$ respectively.}
    \label{fig:vdw}
\end{figure}

We briefly review here some basic thermodynamic ideas in the context of the Van der Waals fluid, and later see how to extend them to fluids with bonding. The Van der Waals fluid is a mean-field model for the liquid-gas phase separation in simple fluids. The Helmoltz free energy per particle $f=F/N$ is

\begin{eqnarray}
\beta f_\text{VdW}&=&\log(\rho)-1  \nonumber \\
&-& \log(1 - b \rho) \nonumber \\
&-&\frac{a \rho}{T}
\end{eqnarray}

where we have decomposed the free energy in its ideal, repulsive, and attractive contributions in each line respectively. 
As  done previously, here and in the following we  assume the thermal length $\lambda=1$. The repulsive contribution is the same (up to the second virial coefficient) as that of hard spheres of volume $b$, and the attractive contribution is a perturbative correction whose  intensity 
is controlled by $a>0$. The Van der Waals model has a critical temperature of $T_c=\frac{8 a}{27 b}$ and a critical volume $v_c=3b$.
The emergence of liquid-gas phase separation at any $T<T_c$ is apparent when plotting the density dependence of the isothermal compressibility $\beta_T=-\frac{1}{V}\frac{\partial V}{\partial P}_{\small{|T}}$, or of its inverse, the isothermal bulk modulus $\kappa_T=1/\beta_T$

\begin{equation}\label{eq:bulkmodulus}
\beta\kappa_T/\rho^2=
\frac{\partial^2\rho\beta f}{\partial\rho^2}
\end{equation}
Thermodynamic stability requires the Helmoltz free energy to be a convex function of the volume, and thus $\kappa_T>0$.

In Fig.~\ref{fig:vdw} we plot the different contributions to the (normalized) bulk modulus for a Van der Waals fluids with $a=0.5$ and $b=0.8$ and for a subcritical temperature of $T=0.15$ (continuous lines). The ideal gas compressibility diverges at low pressures as $\beta_T\sim 1/P$, such that $\beta\kappa_T/\rho^2\sim 1/\rho$, while the excluded volume contribution to the bulk modulus diverges at the density of close packing $\rho_\text{c.p.}=1/b$.
Both terms are purely entropic, and thus their $\beta\kappa_T/\rho^2$ contribution is always positive and does not change with temperature. The attractive term, on the other hand, is of purely energetic origin and negative, and due to this contribution the overall bulk modulus curve (continuous black line) becomes negative in a density interval delimited by the spinodal points (black circles). We notice that the strength of the attractive term becomes more and more negative as temperature is decreased $\beta\kappa^\text{attr}_T/\rho^2=-2 a/T$. This causes the instability region to grow more and more with decreasing temperature, with the behaviour we noted for isotropic potentials in Fig.~\ref{fig:lj}.

\subsection{Wertheim fluid}

\begin{figure}[t!]
    \centering
    \includegraphics[width=1\columnwidth]{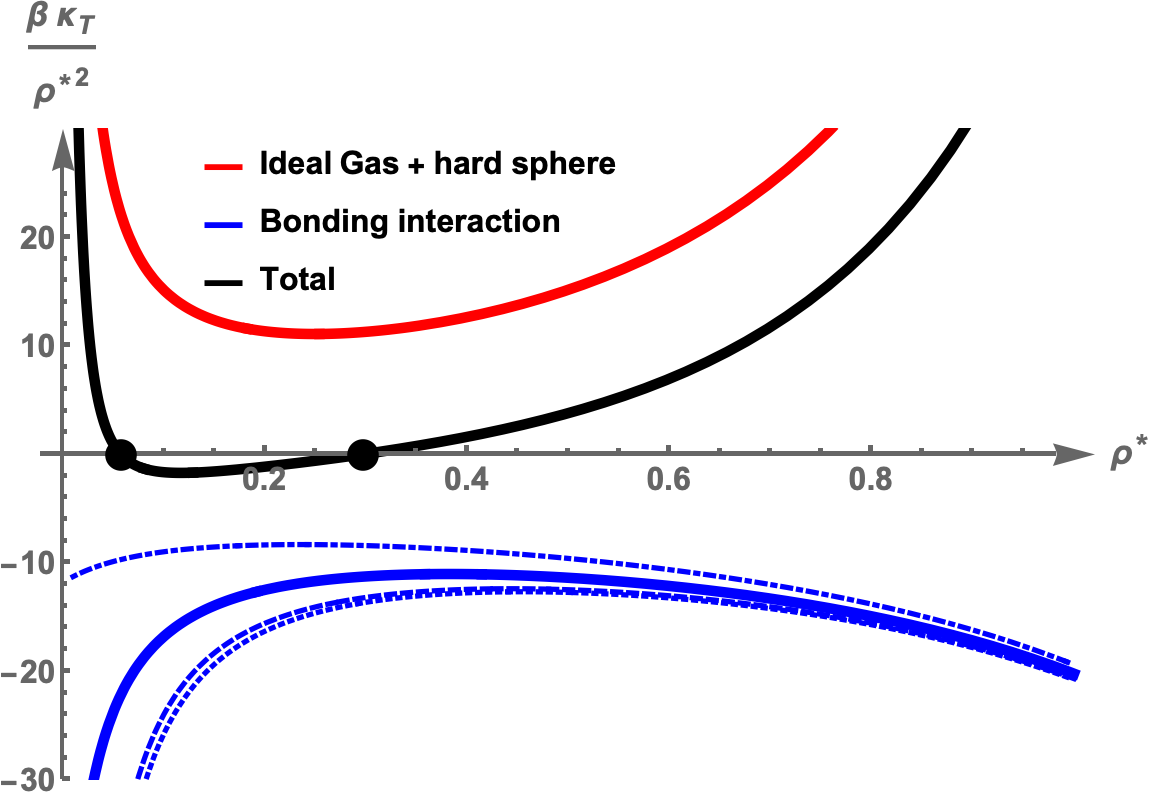}
    \caption{Isothermal bulk modulus, $\beta\kappa_T/\rho^2$, for the Wertheim model with $M=4$ patches, bonding volume $\mathcal{V}_b=0.005$, and $\epsilon=1$ ($T^{*}_c=0.164$). Continuous lines represent the different contributions to the bulk modulus at $T/T_c=0.91$. Full points mark the location of the spinodal points. Dotted, dashed, and dot-dashed lines present the bonding contributions for $T/T_c=0.30$, $T/T_c=0.61$, and $T/T_c=1.22$ respectively.
    }
    \label{fig:compressibiliywertheim}
\end{figure}

The free energy of the Wertheim fluid is also divided into ideal (Eq.~\ref{eqn:fid}) and hard-sphere (Eq.~\ref{eqn:carnahan}) terms that are purely entropic, and a bonding term which has mixed character and is given by Eq.~\ref{eqn:fbond}. The contributions of these terms to the isothermal bulk modulus are plotted in Fig.~\ref{fig:compressibiliywertheim} for a model with $M=4$ patches, bonding volume $V_b=0.005$ and attractive strength $\epsilon=1$.

What differentiates the Wertheim fluid from the isotropic potentials like the Van der Waals fluid is the property of the bonding term. Instead of being constant with density, the bonding term has a maximum towards low values of $\rho$. This causes the bulk modulus to become negative first in this low density region, and it is the reason why phase separation occurs at lower density in patchy models compared to isotropic models. The second important property of Wertheim's bonding term is its low-temperature dependence. Instead of becoming increasingly negative with decreasing temperature, the bonding term converges to a constant profile. This is shown in the blue lines in Fig.~\ref{fig:compressibiliywertheim} which presents the bulk modulus for $T=0.2$ (dot-dashed), $0.15$ (continuous), $0.10$ (dashed), $0.05$ (dotted). The saturation is a consequence of the fact that at low temperatures most bonds are fully formed, and the total energy doesn't change.

These two properties, i.e. the shift to small $\rho$ of the spinodal points and their temperature-independence at low $T$, 
explain the empty liquid behaviour of bonded liquids described before.

Since both isotropic attractive terms and bonding terms give rise to a liquid-gas phase separation, the former at higher densities than the latter, it would be tempting to describe the second liquid-liquid critical point scenario by combining both terms, i.e. by looking at the phase behaviour of patchy particles that also have an isotropic attraction in addition to a patchy contribution, a modification often employed to describe proteins via coarse-grained approaches~\cite{liu2007vapor, gogelein2008simple, fusco2016soft}.
But it is clear from the discussion above that this is not generally possible: the effect of the isotropic term would be to add a constant negative term to  the bulk modulus in Fig.~\ref{fig:compressibiliywertheim}, making the coexistence window larger but it would  be unable to  split the density range  into two disconnected regions.

In the following we will introduce a mechanism that quite generally allows for the formation of a second region of instability (i.e. a liquid-liquid transition) in bonded liquids that is not encompassed by the previously discussed mean-field or Wertheim approaches.

\subsection{Optimal network forming density}\label{sec:optimal_network_density}

The mean-field theory of bonding and the  Wertheim's perturbation approach do not contain information on the geometrical arrangement of patches on the surface of the particle. These approaches give reasonable results when the density is low and the fraction of possible bonds is far from one. But the formation of a tetrahedral network is severely hindered at high densities by geometrical constraints, where packing is not compatible with the tetrahedral bonding geometry. Indeed, increasing the density beyond an optimal value can be achieved only at the expense of  breaking some of the bonds. We thus expect the bonding energy to be a non-monotonous function of density.
We call the density where, at any given temperature, the bonding energy has a minimum the \emph{optimal network forming density}~\cite{de2006dynamics}. At sufficiently low temperatures, the existence of an optimal density promotes phase separation.

We now show that if the optimal density is beyond the liquid-gas coexistence region (which in empty liquids does not extend to high densities) it can give rise to a separate instability region (a liquid-liquid phase separation). We start by taking the expression for the isothermal bulk modulus, Eq.~\ref{eq:bulkmodulus}, and split the free energy in its energy and entropy contributions.
\begin{equation}
\beta\kappa_T/\rho^2=\frac{\partial^2\rho\beta f}{\partial\rho^2}=\beta\frac{\partial^2\rho u}{\partial\rho^2}-\frac{\partial^2\rho s}{\partial\rho^2}
\end{equation}

where $u=U/N$ and $s=S/N$ are the internal energy and entropy per particle. At low temperatures the contribution of the entropic term becomes negligible, and the behaviour of the bulk modulus is dominated by the shape of the energy function~\cite{sciortino1997line}. A localized minimum in the energy function $u=u(\rho)$, such as the one at the optimal network density, can cause system instability ($\kappa_T<0$) at some density $\rho_0$ if

\begin{equation}
\left.\frac{\partial^2\rho u}{\partial\rho^2}\right\rvert_{\rho=\rho_0}=2\left.\frac{\partial u}{\partial\rho}\right\rvert_{\rho_0}+\left.\rho_0\frac{\partial^2 u}{\partial\rho^2}\right\rvert_{\rho_0}<0
\end{equation}

For example, the steep decrease in energy near the minimum $u'(\rho_0)< 0$ or the change of curvature after the minimum $u''(\rho_0)<0$ can destabilize the liquid and promote a second phase separation at sufficiently low temperatures.

Energy minima have been found in a variety of tetrahedral patchy models, including the primitive model for water~\cite{de2006dynamics}, the Kern-Frenkel potential~\cite{saika2013understanding}, and the tetrahedron origami model~\cite{ciarella2016toward}. The energy minima originates from geometric constraints to bonding, and in most models it is indeed found to be temperature independent~\cite{de2006dynamics,saika2013understanding,starr2014crystal,ciarella2016toward}. The same temperature independence of the minimum is found in molecular models of water~\cite{sciortino2011study,poole2011dynamical}.

Patchy models allow us also to study the dependence of the energy minimum on the width of the angular interaction. In particular, in the Kern-Frenkel model of Eq.~\ref{eq:kernfrenkel1}~and~\ref{eq:kernfrenkel2} it is observed that the minimum becomes more and more pronounced as the width of the angular interaction ($\theta_\text{max}$) is reduced~\cite{saika2013understanding}, while also shifting to lower densities.

It was shown~\cite{munao2011simulation,saika2013understanding} that in a variety of tetrahedral model (patchy particles, BKS silica, ST2 and TIP4P/2005 water) the appearance of the minimum in $U(\rho)$ is concurrent with the appearance of a pre-peak in the structure factor $S(q)$. The pre-peak, also called the first diffraction peak (FSDP), appears at wavenumbers $k d/2\pi<1$, i.e. at distances longer than the average nearest-neighbour distance $d$, and is widely-observed in tetrahedral liquids such as $Si$, $Ge$, $SiO_2$, $GeO_2$, $BeF_2$, etc. It was recently demonstrated that the peak originates from the scattering of  tetrahedral units~\cite{shi2019distinct}, and can be decomposed in two populations providing direct evidence for the two-state interpretation of water's anomalies~\cite{shi2020direct} which we will discuss briefly later.

\subsection{Optimal network model}

\begin{figure}[t!]
    \centering
    \includegraphics[width=1\columnwidth]{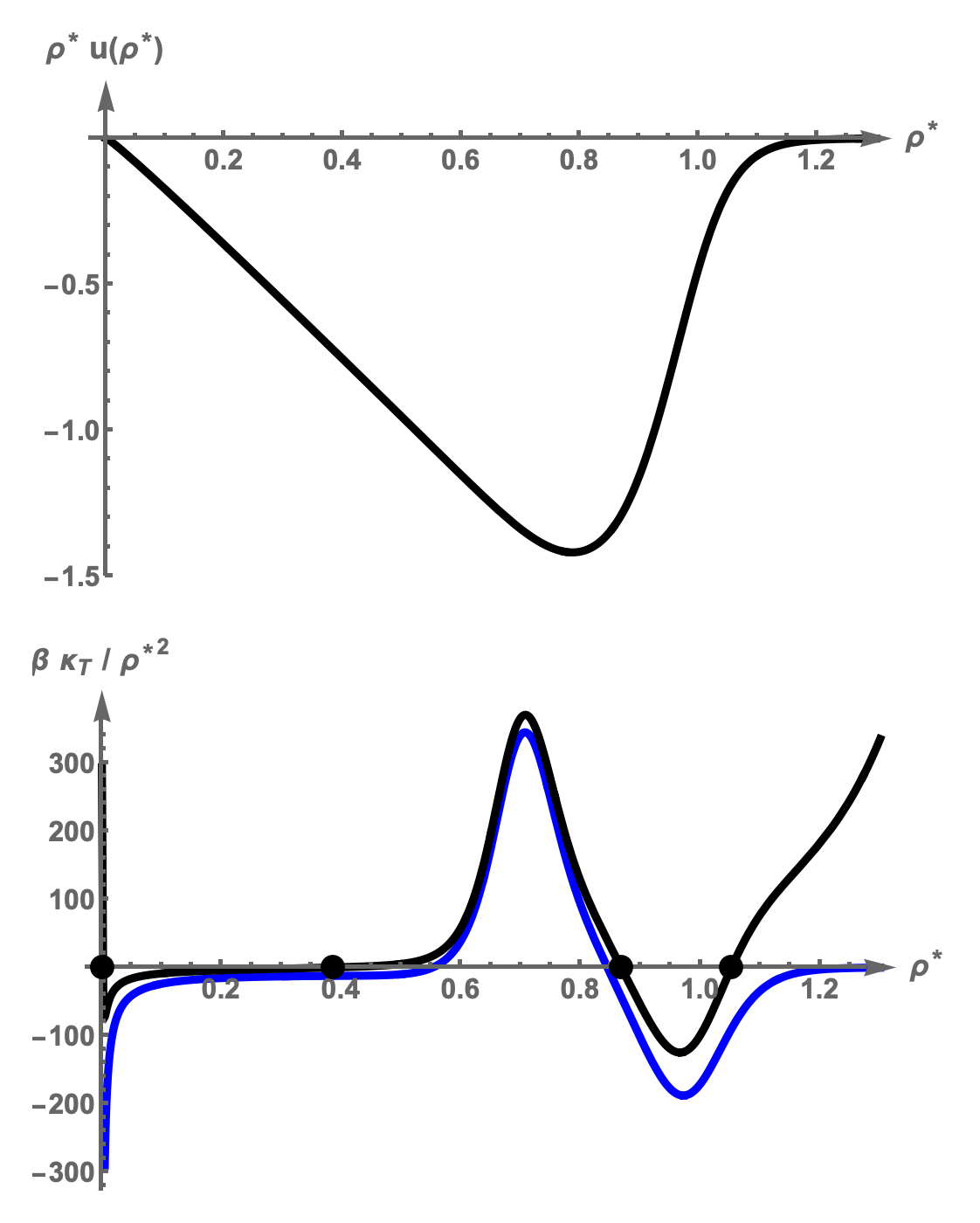}
    \caption{(top panel) Profile of $\rho u$ as a function of $\rho$, displaying the minimum at the optimal density  for the optimal network model with $M=4$ patches, bonding volume $\mathcal{V}_b=0.005$, $\epsilon=1$, $\rho_\text{on}=0.7$, $\sigma_\text{on}=1/30$.
    (bottom panel) Isothermal bulk modulus, $\beta\kappa_T/\rho^2$. The blue line represents the bonding contribution to the bulk modulus at $T^{*}=0.1$, while the black line is the modulus of the full model, including the ideal and hard sphere contributions. Full points mark the location of the spinodal points.
}
    \label{fig:optimalmodel}
\end{figure}

To give a concrete example of the previous discussion, we build here a zeroth-order optimal network model  
by  modulating the bonding volume of Eq.~\ref{eqn:wertheimbonding} with a rapidly decaying function of density

\begin{equation}\label{eqn:optimalmodel}
V_b\rightarrow V_b \left({1-\frac{1}{1+e^{-\frac{\rho-\rho_\text{on}}{\sigma_\text{on}}}}}\right)
\end{equation}
where $\rho_\text{on}$ is the optimal network density and $\sigma_\text{on}$ expresses how steep the bonding is suppressed with increasing $\rho$. The choice of the logistic function in Eq.~\ref{eqn:optimalmodel} is in itself arbitrary, and chosen to qualitatively represent the effects of 
the progressive bond disruption induced by an increase in the packing fraction beyond the optimal packing.

In Fig.~\ref{fig:optimalmodel} we plot results from this model with the following choices: $V_b=0.005$, $\epsilon=1$, $\rho_\text{on}=0.7$, $\sigma_\text{on}=1/30$. The top panel shows the non-monotonous dependence of $\rho u(\rho)$. The bottom panel plots the isothermal bulk modulus $\beta\kappa_T/\rho^2$ as a function of $\rho$ (black line). The bonding term (blue line) is responsible for the loss of stability of the fluid ($\kappa_T<0$) in two distinct regions of density, delimited by spinodal points (full points in the figure). The (empty-)liquid-gas phase  separation occurs at   low values of $\rho$, while a second transition (the liquid-liquid transition) appears at larger values of $\rho$ in the same region where the function $\rho u(\rho)$ (top panel) becomes a concave function. The $\rho u(\rho)$ function displays a minimum near the optimal network forming density $\rho_\text{on}$. The optimal network model connects the presence of a liquid-liquid transition in network models with the presence of an optimal density of assembly, which is itself reflected in a minimum of the energetic bonding terms that contribute to $u(\rho)$.

In order of increasing density, the gas, the low-density liquid, and of the high-density liquid are all delimited by spinodal points. The low-density liquid region is delimited by two spinodal points, and this property is connected with the existence of thermodynamic anomalies.  In fact, at the spinodal points the isothermal compressibility $\beta_T=1/\kappa_T$ diverges, so that between the spinodal points there must be a isothermal compressibility minimum (or a bulk modulus maxima as seen in the bottom panel of Fig.~\ref{fig:optimalmodel}). We have thus seen that a sufficient condition for the compressibility anomaly is the presence of two critical points. We will return in more detail on these aspects in the Section on liquid anomalies.

\subsection{Softness and flexibility}

In this section we review what are the general properties of a pair potential that control the location and strength of the energy minimum and the presence of an optimal network density.

The general dependence of the liquid-liquid critical parameters on potential softness can be summarized as follows~\cite{smallenburg2014erasing}. The critical temperature $T_c$ has a maximum when the typical bonded interparticle distance has a range similar to the hard-core repulsion, while both increasing or decreasing softness from this optimal value lowers $T_c$. The critical density $\rho_c$, instead is seen to be more or less independent of softness. 

The second potential property that controls liquid-liquid phase separation is \emph{bond angular flexibility}. In the Kern-Frenkel model of Eq.~\ref{eq:kernfrenkel2}, for example, it is controlled by the parameter $\cos\theta_\text{max}$: the higher $\theta_\text{max}$, the higher is the flexibility in the network. Several simulations studies have shown~\cite{smallenburg2014erasing,starr2014crystal} that the critical parameters of the liquid-liquid transition have the following parameter dependence: increasing flexibility results in a lowering of $T_c$ and $P_c$ and an increase of the critical density $\rho_c$. As flexibility is increased, the liquid-liquid phase separation can become metastable with respect to gas-liquid phase separation: the intersection of liquid-liquid binodals with the liquid-gas spinodal gives rise to points where one of the two liquids (LDL) becomes metastable to gas cavitation. These special termination points have been named Speedy points~\cite{chitnelawong2019stability}.

Bond flexibility appears also at the level of density fluctuations: it was shown that decreasing bond flexibility is accompanied with the emergence of the first diffraction peak (FSDP), with a concurrent increase in the isothermal bulk modulus of the liquid~\cite{saika2013understanding}. As we discussed before, the FSDP is one of the structural signs of the emergence of a minimum in the energy function $u(\rho)$. These results show that there is a very general coupling between density fluctuations, the bulk modulus of the system, and the emergence of a energy minimum, that gives rise to liquid anomalies and, possibly, to liquid-liquid phase separation.

The same network properties have been considered in water. A complete network inter-penetration is seen in ice VI/XV, which structure is composed of two interpenetrating open zeolite-like frameworks, in ice VII/VIII/X, which consist of two interpenetrating cubic ice lattices~\cite{KuoIceVII2004,salzmann2019advances} and in the trigonal, metastable ice IV~\cite{EngelhardtStructure1981}.
 
The concept of network interpenetration could possibly be also of relevance in the description of the dense  
amorphous form of water, both liquid and glass.
In this case, interpenetration should not be identified as
completely independent bonding networks (as in the crystal phases) but as two networks close-by in space  which are distinct only within a finite distance~\cite{hsu2008hierarchies}.  
The observation that
 HDL has a density $\sim20-25\%$ higher than LDL suggests 
 that some sort of interpenetration could take place.
 The present evidence points in the direction that
 complete interpentration is not found in the high density amorphous ice (HDA) produced from the isothermal compression of hexagonal ice or low density amorphous ice (LDA)~\cite{soper2000structures,shephard2017high,martelli_searching2018} nor in simulated configurations of the HDL. 
 In both HDA and LDA
 an anomalous suppression of long-range density fluctuations~\cite{martelli_hyperuniformity_2017} comparable to that of amorphous silicon~\cite{HejnaNearly2013,XieHyperuniformity2013} has been observed, suggesting a network topology resembling a continuous random network~\cite{WootenComputer1985,BarkemaHigh2000}. 
 The lack of complete interpenetration in HDA has been attributed to the complex kinetics and high energy required to generate an interpenetrating network starting from a non interpenetrating one~\cite{shephard2017high}, and to the loss in configurational entropy~\cite{martelli_searching2018}. On the other hand, clear connections between ice IV and HDA have been reported~\cite{salzmann_2002,martelli_searching2018}, hence suggesting that the observation of partial interpenetrating network in HDL, while elusive, can not be discarded. As a matter of fact, the network topology of liquid water has been inspected at both ambient conditions~\cite{SantraLocal2014,MartelliTopology2021} and upon approaching the LLCP~\cite{palmer2014metastable,martelli_unravelling_2019,ShiNetwork2019,martelli_rings}, but the degree of interpenetration has not been investigated.

From the discussions above, we could conclude that the best conditions for the observation of a liquid-liquid transition are obtained with potentials that have a low amount of flexibility. But there is one aspect that we haven't touched so far, and it's the interference of the liquid-crystal transition. We will show that flexibility is also the key factor controlling the crystal-forming ability of reduced valence models, which affects indirectly the possibility of observing a liquid-liquid transition.

\section{Liquid anomalies}

By liquid anomalies we refer to thermodynamic anomalies, i.e. the anomalous behaviour of the thermodynamic response functions of a fluid that show  a non-monotonous change when varying some thermodynamic parameter. The location of maxima or minima in a response function can be used to trace thermodynamic lines on the phase diagram~\cite{sastry1996singularity,rebelo1998singularity,poole2005density,lynden2005computational,martelli_unravelling_2019}.
The schematic phase diagram in Fig.~\ref{fig:llcp}, encoding the second critical point scenario, shows  the anomalous lines that are most commonly considered~\cite{poole2005density}

\begin{itemize}
\item $\tilde\rho$ defined as the line where the thermal expansion coefficient 
$\alpha=-\left.\frac{1}{V}\frac{\partial V}{\partial T}\right\rvert_{P}=0$.
This line thus corresponds to a line where density has an extremum (i.e. a density maxima or minima {\it along isobars} as a function of $T$);
\item $\tilde\beta_T$ defined as the extremum line of the isothermal compressibility {\it along isobars} $\beta_T$, where $\left.\frac{\partial\beta_T}{\partial T}
\right\rvert_{P}=0$
\item $\tilde c_P$ defined as the extremum line of the specific heat $c_P$ {\it along isotherms}, where $\left.\frac{\partial c_P}{\partial P}\right\rvert_{T}=0$
\end{itemize}

Anomalies are usually (but not necessarily) associated with the presence of a critical point, from which lines of maximum compressibility and specific heat emanate~\cite{xu2005relation}. Moreover, in the second liquid critical point scenario these lines have to necessarily become lines of minimum for the same response functions. Mean-field spinodals are lines where both compressibility and specific heat diverge

\begin{align*}
\beta_T&\rightarrow\infty \\
c_P&\rightarrow\infty 
\end{align*}

Since the LDL liquid is bounded at low and high $\rho$ by two spinodals, it has to develop a
specific heat minimum (which corresponds to the $\tilde c_P$ line in Fig.~\ref{fig:optimalmodel}) ) at intermediate densities, along an isotherm. 
Similarly the compressibility also diverges along both spinodals.

The $\tilde c_P$ line is also linked to the $\tilde\rho$ line by the following thermodynamic relation~\cite{poole2005density}

\begin{equation}
\left.\frac{\partial c_P}{\partial P}\right\rvert_{T}=-T \left.\frac{\partial^2 v}{\partial T^2}\right\rvert_{P}
\end{equation}

$\tilde c_P$ is thus also the line where the density changes concavity with temperature, and the intersection between the $\tilde c_P$ and $\tilde\rho$ lines is where the density maxima becomes a density minima, as shown schematically in Fig.~\ref{fig:llcp}.

In summary, liquid anomalies are strongly connected~\cite{sastry1996singularity, rebelo1998singularity, poole2005density,holten2017compressibility,goy2018shrinking,caupin2019thermodynamics, martelli_unravelling_2019,fijan2020progression,tanaka2020liquid}
 and the presence of a second liquid critical point is a sufficient but not necessary condition for their appearance. The anomalies can persist also in the limit in which the liquid-liquid critical temperature goes to zero. It is interesting to note that \emph{remnants} of criticality in absence of a critical point have recently been discovered also for the liquid-gas critical point in patchy particles in the limit where the valence goes to two ($M\rightarrow 2$)~\cite{stopper2020remnants} and patchy models with different patch types~\cite{Tavares_2020}.

\subsection{Optimal networks and anomalies}

We now look for a connection between the existence of an 
optimal networks density (a non-monotonous energy function $u(\rho)$) and the presence of anomalies. Starting from the thermodynamic definition of pressure

\begin{equation}
\left.\frac{\partial f}{\partial v}\right\rvert_{T}=-P
\end{equation}
we write the free energy per particle as $f=u-T s$, and use the Maxwell relation $\left.\frac{\partial s}{\partial v}\right\rvert_{T}=\left.\frac{\partial P}{\partial T}\right\rvert_{V}$ to rewrite the previous equation as

\begin{equation}
\left.\frac{\partial u}{\partial v}\right\rvert_{T}=T\left.\frac{\partial P}{\partial T}\right\rvert_{V}-P
\end{equation}

Re-writing the left hand side as a derivative with respect to density, $\rho=1/v$, and remembering that 
$\left.\frac{\partial P}{\partial T}\right\rvert_{V} = -
\left.\frac{\partial V}{\partial T}\right\rvert_{P}  
\left.\frac{\partial P}{\partial V}\right\rvert_{T} 
$ 
the first term on the right hand side can be identified with the thermal pressure $\partial P/\partial T=\alpha\kappa_T$

\begin{equation}
\rho^2\left.\frac{\partial u}{\partial\rho}\right\rvert_{T}=-T\alpha\kappa_T+P
\end{equation}
At each temperature, density extrema are identified by the condition that the thermal expansion coefficient is zero ($\alpha=0$), so a density anomaly arises if a density value $\rho_{ext}$ exists for which  $\rho^2\left.\frac{\partial u}{\partial\rho}\right\rvert_{T}$ becomes
identical to $P$, i.e.

\begin{equation}\label{eq:densityanomaly}
\rho^2\left.\frac{\partial u}{\partial\rho}\right\rvert_{T}(\rho_{ext})=P(\rho_{ext})
\end{equation}

The existence of a line of density anomalies at positive pressure then
requires that $\partial u/\partial\rho>0$ at $\rho_{ext}$, i.e.  a non-monotonic dependence
of the energy vs density curve.  Note that for all models a region of $\partial u/\partial\rho>0$ is expected to exist, when density has 
increased so much that the repulsive part of the potential has become the
dominant contribution to $u$.  But the region of density where  $\partial u/\partial\rho>0$ is commonly embedded inside the glass (or crystal) regions.
In the case of empty liquids, $\partial u/\partial\rho>0$ may instead appear for a density $\rho_{ext}$
well before dynamic arrest takes place. 
 Because $\alpha>0$ in a normal liquid, this condition corresponds to the density maximum anomaly. It is an observable density maximum as long as the $(\rho,T)$ location of the maximum is outside the spinodal region, e.g. for temperatures above the critical point. Eq.~\ref{eq:densityanomaly} can also be satisfied at negative pressures, with the solution usually corresponding to a density minimum. The possibility of measuring a density minimum is subject to the condition that it is not pre-empted by gas cavitation or crystallization of the liquid.

\begin{figure}[t!]
    \centering
    \includegraphics[width=1\columnwidth]{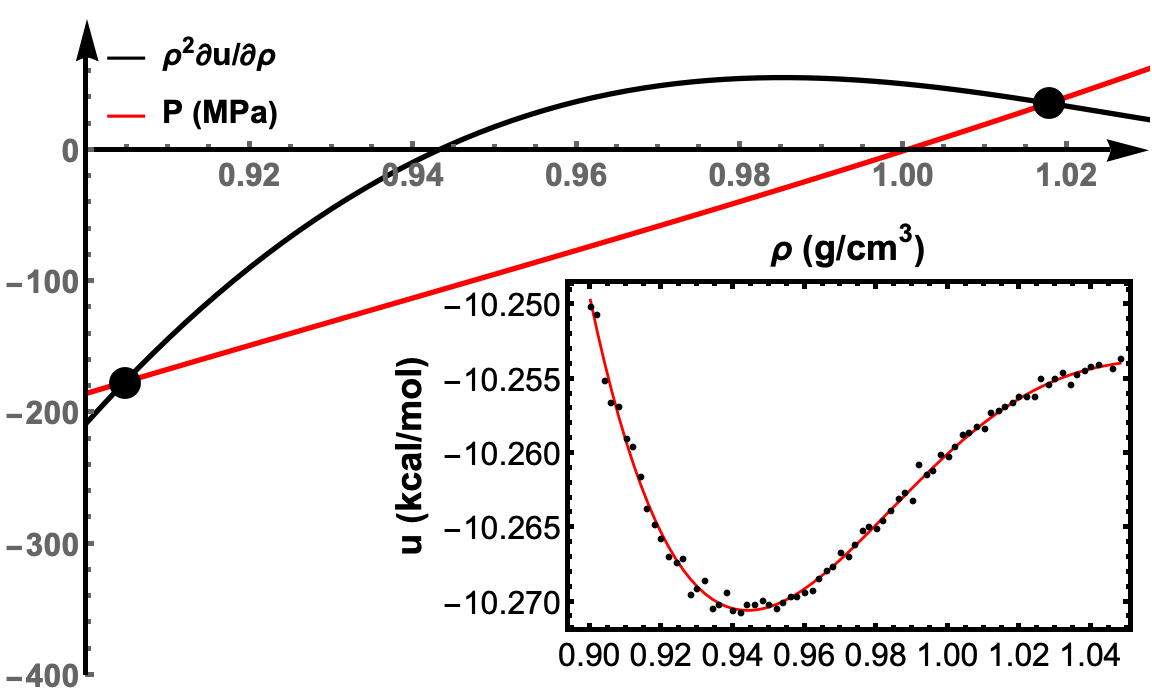}
    \caption{Relation between the energy minimum and the density anomaly. The inset shows the energy computed for a system of $N=4000$ TIP4P/2005 molecules in the NVT ensemble at $T=270$~K. Each point is the average over $5$~ns simulation runs, while the line is the fit with a fourth order polynomial. The main panel displays the two terms in Eq.~\ref{eq:densityanomaly} which intersect at the location of the density maximum at $P=36\pm5$~MPa and the density minimum at $P=-177\pm5$~MPa, both plotted with points.}
    \label{fig:tip4p}
\end{figure}

To illustrate the connection between the energy minimum and the density anomaly we run simulations of the TIP4P/2005 model of water~\cite{abascal_general_2005,gonzalez2016comprehensive} and show the results in Fig.~\ref{fig:tip4p}. At $T=270$\,K the model displays an energy minimum (in the inset) which gives rise to both a density minimum and a density maximum within the region of (meta-)stability of the liquid, that correspond to the intersection between the red and black lines at negative and positive pressures respectively.

\subsection{Alternative Models}

So far we have entirely focused on the thermodynamic properties of water from the perspective of network fluids, and in particular of tetrahedral networks. Alternative models of the liquid-liquid transition and of thermodynamic anomalies have been built from different, and sometimes complementary, perspectives. Without any pretense to being exhaustive, we briefly summarize some of the most popular approaches.

It has long been known that water-like anomalies also appear in particles interacting with isotropic potentials that have two lengthscales, comprising a short-range hard-core repulsion and an intermediate range soft-core repulsion~\cite{mishima1998relationship,Tu_2009,Tu_2012,ryzhov2020complex}. In these models two typical inter-particle distances compete at different thermodynamic conditions. The first models with these properties were the repulsive-shoulder models of Hemmer and Stell \cite{hemmer1970}, and anomalous properties where then extended to potentials with similar characteristics, like ramp potentials~\cite{kumar2005static}, square-shoulder potentials~\cite{franzese2001generic,Skibinsky_2004,malescio2005liquid,de_Oliveira_2006,gribova2009waterlike} and its continuous version \cite{franzese2007differences}, the Jagla potential~\cite{jagla1999core,xu2006thermodynamics}, etc.
Another of such models \cite{debenedetti1991}, where the short length-scale is attractive and the long one is repulsive, still has anomalies and can form a low density open structure at low pressure and temperature.
In water, the short and long lengthscales are associated respectively with the isotropic repulsion and the hydrogen bond length. The isotropic two-lengthscale potentials are thus thought as orientationally averaged models of the orientation-dependent interaction of water. 
Indeed, by solving the Ornstein-Zernike integral equation, it has been found \cite{head-gordon1993} that the isotropic potential corresponding to the experimental oxygen-oxygen (averaged over orientational degrees of freedom) pair correlation function displays a core-softening similar to the continuous version of Hemmer and Stell models.
The anomalies shown by core-softened models depend on the details of the potential and in general their presence is neither a sufficient nor a necessary condition for the appearance of a liquid-liquid transition. Even though these models have water-like anomalies, and in some case with the same hierarchical order of water, and can show a LL critical point, the entropic behavior they predict displays a slope in the P-T phase diagram opposite to that shown by water. Beyond the description of some property of water, these models have found application in the study of solutions of globular proteins, colloids, and liquid metals.
The definition of isotropic potentials naturally lends itself to the use of thermodynamic theories, such as the modified van der Waals theory, and perturbation theories~\cite{fomin2006generalized}.

An intermediate approach between the isotropic potentials and the network-based approach is given by the Hydrogen-bond model, where a bonding free energy is added to the var der Waals free energy~\cite{poole1994effect,chitnelawong2019stability}. The bonding term is built as a two-level system and, crucially, it has a strong density dependence, which allows it to develop a $u(\rho)$ minimum. The hydrogen bond model has been successful at unifying the different thermodynamic scenarios for water, and in particular the transition between the stability-limit scenario and the two-critical points scenario. A first-principle expression of the bonding term has also been derived~\cite{truskett1999single}.

A pivotal role in the understanding of the different water scenarios has been played by lattice models. We mention here the Sastry's model~
\cite{sastry1996singularity}, extended by  Franzese and Stanley~\cite{Franzese_2002} and fully characterized in later works \cite{stokely2010effect,coronas2020franzese,bianco2014,bianco2019} to naturally incorporate the hydrogen bond cooperativity as a tuning parameter in the model.

A different and more microscopic approach to the description of water anomalies is taken by \emph{two-state models}. The focus of these models is on the structural change that occurs locally to the network of hydrogen-bonded molecules~\cite{moynihan1996two,tanaka2000simple,cuthbertson2011mixturelike,nilsson2011perspective,tanaka2012bond,holten2012entropy,shi2018common,camisasca2019proposal,martelli_unravelling_2019}. The environment surrounding a water molecule is divided in two populations: an \emph{ordered} population of energetically favoured environments, and a population of entropically favored environments.
Evidence for this bimodality has been corroborated by analysis of numerical~\cite{russo2014understanding,shi2018common,martelli_unravelling_2019,montes2020structural} and experimental data~\cite{nilsson2012fluctuations,taschin2013evidence,gallo2016water,shi2020direct,woutersen2018liquid,pettersson2019x,camisasca2019proposal,kringle2020reversible}.
The two states inter-convert into each other in equilibrium, and their relative composition depends on the thermodynamic state point: the ordered state being the (free energy) favored state at low pressures and temperatures, and the disordered state being the majority component at high temperatures and/or pressures.
The advantage of these models is that the free energy (which is taken as the one for a regular mixture) can be expressed directly in terms of the compositions extracted from numerical simulations~\cite{russo2014understanding,shi2018common} or experimental data~\cite{shi2020direct,shi2020anomalies}, predicting rather than fitting the anomalous behaviour. A general theory of polyamorphism, not only related to water, has been introduced in the context of two-state models~\cite{anisimov2018thermodynamics}.
For a detailed introduction to two-state models and their applications we refer to the recent review in Ref.~\cite{tanaka2020liquid}. Last but not least, for a model which combines the physics of mixing and Wertheim's perturbation theory see Ref.~\cite{smallenburg2015liquid}.

\section{Crystallization}
Tetrahedral systems exhibit puzzling crystallization behaviour. For example, water and silica both form tetrahedral networks, but the first one is a crystal-former while the second one is the prototypical glass former.
Even before being applied to the liquid-liquid transition in water, tetrahedral patchy models were popular models to study the crystallization of limited-valence colloidal particles. A lot of attention was put in understanding the conditions for successful nucleation of open crystalline structures, with particular emphasis on the diamond cubic crystal. In this section we briefly review what are the most important geometric factors that control the crystallization of patchy particles, and we then highlight how these affect our ability to observe liquid-liquid transitions.

\subsection{Phase diagram}

The crystalline structures that are most prominent in the phase diagram of tetrahedral patchy particles are:

\begin{description}
\item[diamond cubic (dc)] it is isostructural to cubic ice, ice I$_c$. In the \emph{dc} structure the bond angles (angles formed by triplets of nearest neighbours) are perfectly tetrahedral ($\sim109.5^\circ$), which allows this structure to be stable even at very low patch widths.
\item[hexagonal diamond (dh)] it is isostructural to hexagonal ice, ice I$_h$. It is a polytype of the \emph{dc} structure, meaning that the two crystal can stack along the basal plane of the \emph{dh} structure with very low free energy penalty. From a free-energy standpoint, the \emph{dh} structure is virtually indistinguishable from the \emph{dc} structure, and nucleation always results in a mixture of both polytypes. For brevity in the following we will refer exclusively to the \emph{dc} phase. Avoiding stacking faults to obtain a pure colloidal \emph{dc} crystal is currently one of the major challenges of nanotechnology~\cite{romano2020designing}.
\item[body-centered cubic (bcc)] it is isostructural to ice VII/VIII/X. The \emph{bcc} structure can be seen as the sum of two interpenetrating \emph{dc} lattices. For this reason, its bonding angles are tetrahedral, and the density is roughly twice as that of the \emph{dc} crystal.
\item[face-centered cubic (fcc)] it is isostructural to ice XVIII~\cite{millot2019}. Bond angles in the fcc crystal are not perfectly tetrahedral, but bonds are possible if patches are wide enough. Also an orientationally disorder \emph{fcc} is often present, where the patchy particles occupy the position of the \emph{fcc} lattice, but with random orientations.
\end{description}

At small pressures $P$, the dominant phases are the \emph{dc} and the \emph{dh} phases, because of their low density and high entropy. The \emph{dc} phase is stable only around a small range of densities, which is usually close to the energy $u(\rho$) minima~\cite{saika2013understanding}, confirming the idea that the minimum is related with the establishment of a tetrahedral network. The small compressibility of the \emph{dc} phase compared to the liquid phase is also responsible for re-entrant melting at high densities in some models~\cite{romano2010phase}.
The \emph{bcc} phase is found at intermediate pressures, while the high pressure phase diagram is where the dense \emph{fcc} phase is stable (with an orientationally disordered \emph{fcc} at high $T$). 

\subsection{Crystal forming ability}

\begin{figure}[t!]
    \centering
    \includegraphics[width=1\columnwidth]{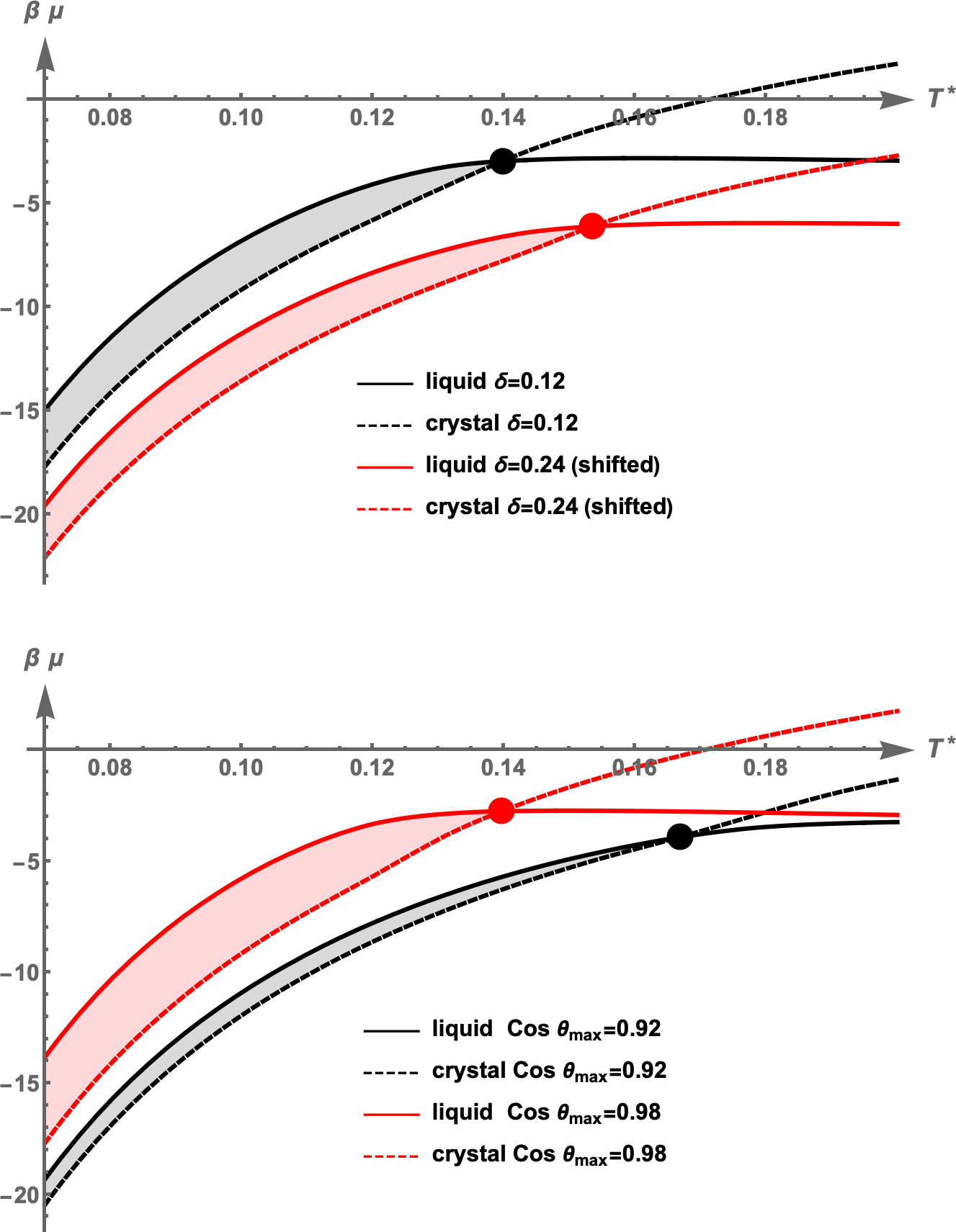}
    \caption{Chemical potentials $\beta\mu$ for liquids (from Wertheim theory, solid lines) and crystals (from Eq.~\ref{eq:solid}, dashed lines) at different values of the parameters $\delta$ and $\cos(\theta_\text{max})$. The pressure is set to $\beta P\sigma^3=0.01$ (to avoid possible interference from liquid-gas phase separation). Driving forces (chemical potential differences between liquid and crystal below the melting point) are represented by shaded regions. (top panel) Comparison between $\delta=0.12$ (black curves) and $\delta=0.24$ (red curves) at fixed  $\cos(\theta_\text{max})=0.96$. (bottom panel) Comparison between $\cos(\theta_\text{max})=0.92$ (black curves) and $\cos(\theta_\text{max})=0.98$ (red curves) at fixed $\delta=0.24$. }
    \label{fig:crystal}
\end{figure}

We now consider the crystallization of open-crystalline structures, the \emph{dc} and \emph{dh} crystals, as they are the most relevant for water crystallization at ordinary pressures. We will focus on the Kern-Frenkel model (Eq.~\ref{eq:kernfrenkel1}-\ref{eq:kernfrenkel2}), as it allows to consider independently the effects of the radial extent of the attraction (through the parameter $\delta$) and the effects of the width of the angular interaction (through the parameter $\theta_\text{max}$). Open crystalline structures are stabilized by entropy~\cite{romano2010phase, mao2013entropy, sciortino2019entropy}, and in the Kern Frenkel model the bonding entropy can be estimated with simple geometric considerations.

We start with the partition function $Q$ of a particle in the solid as described by cell theory~\cite{sear1999stability}

\begin{equation}\label{eq:partitionfunction}
Q=v_f e^{\beta z\epsilon/2}
\end{equation}

where $z=4$ is the number of bonded neighbours in the \emph{dc} crystal, $\epsilon$ is the interaction strength, and $v_f$ is the free volume that a patchy particle can explore while being bonded in the crystalline configuration. Thanks to the decoupling of translational and rotational degrees of freedom of the Kern-Frenkel model, $v_f$ can be written as the product of translational $v_f^t$ and orientational $v_f^r$ contributions. The rotational term $v_f^r$ has the same scaling as found in the bonding volume (Eq.~\ref{eqn:bondingvolume})

\begin{equation}
v_f^r\propto (1-\cos(\theta))^2
\end{equation}

In a close packed structure, like the \emph{fcc} crystal, the translational volume would scale as the cube of the interaction range $\delta$, but in the Kern-Frenkel model for open crystalline structures it is instead found that it scales as~\cite{romano2010phase}

\begin{equation}
v_f^t\propto\delta^2
\end{equation}

Substituting these expressions in Eq.~\ref{eq:partitionfunction} the  expression for the chemical potential of the \emph{dc} (or equivalently \emph{dh}) crystal
can be expressed as

\begin{equation}\label{eq:solid}
\beta\mu=-2\beta\epsilon-2\log\left((1-\cos(\theta_\text{max}))\delta\right)+\frac{\beta P}{\rho}
\end{equation}

While only approximate, these expressions allow for an understanding of the parameter dependence of the liquid-solid transition in tetrahedral patchy models. In Fig.~\ref{fig:crystal} we plot the chemical potential $\beta\mu$ for both the liquid and crystal phases as a function of temperature for different choices of the parameters $\delta$ (top panel) and $\cos(\theta_\text{max})$ (bottom panel). We see that the melting temperature (located at the crossing between the liquid and crystal curves) is pushed at lower $T$ by decreasing $\delta$ and/or increasing the bond angle $\theta_\text{max}$.  This trend is general in patchy models and is also true for the high pressure phases: the liquid phase is stabilized with a shorter interaction range and/or a wider bond angle. What distinguishes the two parameters $\delta$ and $\theta_\text{max}$ is their role in controlling the crystal-forming ability of the system, i.e. on the temperature dependence of the nucleation rate~\cite{romano2009role,romano2010phase}. To illustrate this point we focus here on the thermodynamic driving force, defined as the chemical potential difference $\beta\Delta\mu$ between the crystal and liquid phase, which is one of the parameters that enters the crystal-forming ability of the system~\cite{russo2018glass}. In Fig.~\ref{fig:crystal} we represent $\beta\Delta\mu$ with the shaded region between the liquid and crystalline curve below the melting temperature. In the top panel we can see that changing $\delta$ at constant $\cos(\theta_\text{max})$ doesn't significantly alter the driving force for nucleation, while in the bottom panel we note instead that the driving force increases more substantially with decreasing $T$ for models with low bond angle (high $\cos(\theta_\text{max})$). This observation holds generally: the smaller the bond angle $\theta_\text{max}$ the largest is the driving force for nucleation and the crystal-forming ability.
Despite having a lower melting temperature, models with high values of  $\cos(\theta_\text{max})$ are thus the best crystal-formers. It was indeed by reducing the bond angle that the first atomistic simulations of spontaneous \emph{dc} crystal nucleation were observed~\cite{zhang2005self,romano2011crystallization}.

\begin{figure}[t!]
    \centering
    \includegraphics[width=1\columnwidth]{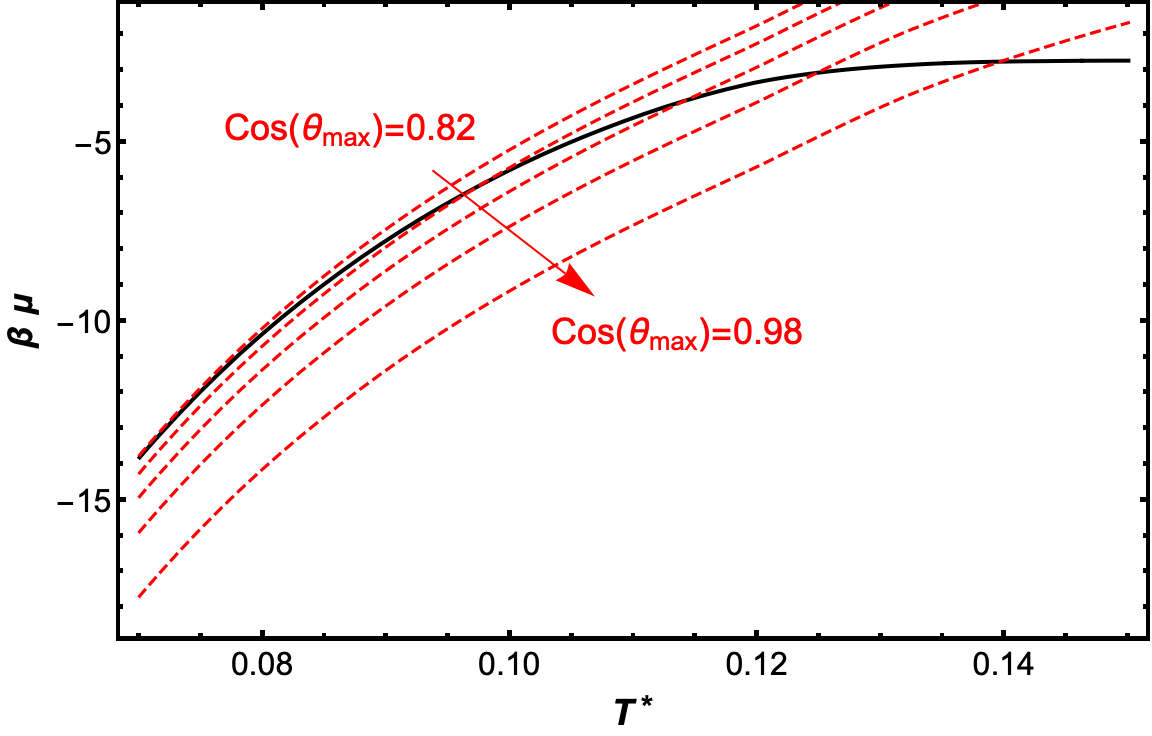}
    \caption{Chemical potentials $\beta\mu$ for liquids and solids for a fixed bonding volume $\mathcal{V}_b=0.00038$ and for different values of the parameters $\delta$ and $\cos(\theta_\text{max})$. The liquid is plotted as a black continuous line, which does not depend on the choice of $\cos(\theta_\text{max})$. Crystal phases are plotted with dashed red lines and range from $\cos(\theta_\text{max})=0.82$ to $\cos(\theta_\text{max})=0.98$ with $\Delta\cos(\theta_\text{max})=0.04$ intervals.}
    \label{fig:ultrastableliquids}
\end{figure}

In Fig.~\ref{fig:ultrastableliquids} we plot the chemical potential $\beta\mu$ for models having the same value of the bonding volume $\mathcal V_\text{b}$ but different value of the paramers $(\delta,\cos(\theta_\text{max}))$. Since in Wertheim theory the free energy depends only on the bonding volume, the chemical potential curve of the liquid phase is the same for all these models, and is plotted as a continuous black line in Fig.~\ref{fig:ultrastableliquids}.
For the crystals the chemical potential (Eq.~\ref{eq:solid}) depends on both  $\delta$ and $\cos(\theta_\text{max})$, and curves for different values of these parameters are represented by red dashed lines.
The slope of these line at different T is given by
\begin{equation}
\left.\frac{\partial\beta\mu}{\partial T}\right\rvert_P=-\frac{\beta h}{T}
\end{equation}
where $h=u+P/\rho$ is the enthalpy per particle. At low temperatures the liquid forms a fully bonded network with an enthalpy very similar to that of the solid: i.e. the liquid and solid lines have similar slopes. With increasing $T$ the liquid curve becomes progressively less steep as the number of broken bonds increases, eventually becoming flat at high $T$. Crystal-forming ability is thus obtained only if the melting temperature (crossing between liquid and solid line) occurs in the region where the liquid has a significant fraction of defects, i.e. for small values of $\theta_\text{max}$.
As we increase the bond angle $\theta_\text{max}$ the crossing of the two curves moves in the region where the liquid has few defects, and in which the slope of the liquid and crystal lines is almost identical, meaning that there will be no thermodynamic driving force with decreasing temperature below freezing.

These conclusions have been confirmed with a full analysis of the nucleation barriers in tetrahedral Kern-Frenkel models~\cite{saika2011nucleation}. With decreasing the bond angle (increasing $\cos(\theta_\text{max})$) it is observed that both the thermodynamic driving force $\beta\Delta\mu$ and the free energy barrier for nucleation increase more rapidly with supercooling. At large bond angles, the barriers have a very weak $T$ dependence and are always very large (beyond $50$~k$_B$T). The single-particle dynamics increases with $\cos(\theta_\text{max})$ since there are more defects in the network structure. Interestingly, the surface tension $\gamma$ also increases with $\cos(\theta_\text{max})$, but it is not enough to compensante the effect of the increase in driving force. The competition between the surface tension penalty for nucleation and the driving force produces a non-monotonic shape of the nucleation rate dependence on $\cos(\theta_\text{max})$, with a maximum that is close to $\cos(\theta_\text{max})=0.96$.

The strong dependence of the crystal-forming ability of patchy particles on the bond angle parameter helps explaining the crystal-forming behaviour of molecular models of water. It has long been known that nucleation in molecular models like TIP4P/2005 and ST2 is strongly suppressed compared to the popular mW model (a coarse-grained model with three-body interactions). This difference in crystal-forming ability is usually explained by the lack of orientational degrees of freedom of the mW model. But a comparative study of the structural properties of these water models~\cite{saika2013understanding} has also revealed that molecular models like TIP4P/2005 and ST2 are much more \emph{flexible} and with a lower number of defects compared to monoatomic models (mW and Stillinger-Weber model \cite{stillinger_1985}). The increased structural rigidity and the higher number of defects of the mW model helps explaining why it crystallizes with ease. A linear relationship between the variance of bond angle distributions and the intensity of the first diffraction peak FSDP can also be used to quantify bond flexibility from scattering data.

Another interesting observation is that at very narrow patch width tetrahedral patchy particles often nucleate clathrate structures~\cite{noya2019assembly}. While not being thermodynamically stable with respect to the \emph{dc} phase, the nucleation of clathrate phases is favored kinetically by the large fraction of five-membered rings present in these liquids.

\subsection{Ultra-stable liquids}

The origin of the stabilization of the liquid phase with respect to the crystal with increasing $\theta_\text{max}$ that we observed in Fig.~\ref{fig:crystal} is due to the larger configuration entropy of the liquid state: a fully connected liquid has the same energy of the crystal, but a much larger degeneracy of configurations, i.e. a larger configurational entropy.
In Fig.~\ref{fig:ultrastableliquids} we notice that the driving force becomes smaller and smaller with increasing $\theta_\text{max}$, and eventually it becomes negative, meaning that the liquid is always more stable than the crystal.
This simple model thus predicts a scenario where the liquid is always more stable than the crystal. It is important to note that the Wertheim theory with which the liquid curve has been computed in Fig.~\ref{fig:ultrastableliquids} requires that a single patch cannot form multiple bonds, a condition which is likely to be violated as $\theta_\text{max}$ is increased. To bypass this limitation, modified Kern-Frenkel potentials have been devised that strictly enforce the one-bond-per-patch condition regardless of the width of the angular interactions~\cite{smallenburg2013liquids}. In these models it is possible to reproduce the behaviour of Fig.~\ref{fig:ultrastableliquids}, i.e. changing the melting line at constant bonding volume, without having to change both $\delta$ and $\theta_\text{max}$. The phase diagram of these models has confirmed that a region where the liquid is stable down to $T\rightarrow 0$ opens up in the phase diagram when the width of the interaction is between $\cos(\theta_\text{max})=0.8$ and $\cos(\theta_\text{max})=0.9$, as we also see in Fig.~\ref{fig:ultrastableliquids}.

While not directly relevant for water, these ultra-stable liquids have been experimentally realized in aqueous solutions of DNA-tetramers~\cite{biffi2013phase,rovigatti2014gels,bomboi2016re}.

\subsection{Crystal-clear liquid-liquid transitions}
In Fig.~\ref{fig:ultrastableliquids} we have seen that the melting line at constant bonding volume $V_b$ shifts to lower $T$ when widening the patch angular interaction. As the patches become wider and the disordered liquid phase becomes relatively more stable, it could be in principle possible to observe the liquid-liquid transition without interference from crystallization. This observation has played a fundamental role in the understanding of water's behaviour where a liquid-liquid transition was observed in the ST2~\cite{palmer2014metastable} and in the TIP4P/2005 and TIP4P-Ice~\cite{debenedetti2020second} models of water. The major criticism moved against the liquid-liquid critical point scenario was in fact the possibility that previous observation of the transition were not due to genuine amorphous density fluctuations, but instead due to undetected crystalline clusters~\cite{limmer2011putative}. To resolve this issue (which was later found to be due to an imprecise sampling of the rotational degrees of freedom in the liquid phase~\cite{palmer2018comment}), the temperature dependence of both the liquid-liquid transition and the solid-liquid transition with varying the bond angle parameter $\theta_\text{max}$ at constant bonding volume was considered in a patchy tetramer model~\cite{smallenburg2014erasing}. The model consists of a central particle with four flexible arms carrying the patch interaction. The length of the arms can be tuned to introduce the necessary softness (i.e. network inter-penetrability) to stabilize the liquid-liquid transition at high $T$ (as we discussed previously). With this model it was found that increasing the arm flexibility (again expressed by the parameter $\cos\theta_\text{max}$) suppresses both the melting and the liquid-liquid transition temperatures, but, crucially, the first one decreases much more rapidly than the second.
So while the liquid-liquid transition is metastable with respect to crystallization at large $\cos\theta_\text{max}$, as observed in most models of water, the opposite is true at lower values of $\cos\theta_\text{max}$. In this regime it is possible to observe a genuine liquid-liquid transition without any interference from crystallization. The same considerations were demonstrated on the molecular model of water (ST2) for which the liquid-liquid transition was first discovered~\cite{smallenburg2015tuning}: by increasing the hydrogen-bond angular flexibility it was shown that the previously hypothesized liquid-liquid transition is continuously connected to a stable liquid-liquid transition at temperatures above the melting temperature.
Stable liquid-liquid transitions have been confirmed in different models, like DNA decorated nanoparticles~\cite{starr2014crystal} and patchy origami tetrahedra~\cite{ciarella2016toward}.

\section{Supercooled dynamics}

On top of its usual thermodynamic anomalies, water displays also dynamic anomalies~\cite{de2018fragile}, such as an increase in diffusivity with increasing pressure. These anomalies are linked to the structural properties of the underlying network, but differently from hard-sphere type liquids, the relaxation processes are not dominated by excluded volume effects (i.e. caging) but by bonding~\cite{ranieri2016dynamical,russo2010self,smallenburg2013liquids,kikutsuji2018hydrogen}. Patchy particles are an ideal model to study network glasses because, with appropriately chosen parameters, the liquid phase can be cooled down to very low temperature without interference from phase separation (\emph{empty liquids}) or crystallization (\emph{ultra-stable liquids}). Specialized computational-algorithms have also been devised that allow to considerably speed-up equilibration in the low-$T$ regime (for a recent review see Ref.~\cite{rovigatti2018simulate}).

The dynamical behaviour of tetrahedral networks changes with temperature, as the system goes from an ideal gas of clusters at high temperatures to a fully bonded tetrahedral network at low temperatures. These two regimes are separated by \emph{percolation}, which is the establishment of a spanning network in the system. Several studies~\cite{de2006dynamics,rovigatti2011self,roldan2017connectivity} have shown that in both these regimes the decay of the diffusion coefficient $D$ with decreasing temperature follows an Arrhenius behaviour

\begin{equation}
D\sim\exp\left(-\frac{E_a}{T}\right)
\end{equation}

where $E_a$ is the activation energy that is related to the microscopic process that controls diffusion. Supercooled liquids that follow the Arrhenius behaviour are also called \emph{strong} glass formers. The activation energy of the high-T regime is smaller than the one for the fully bonded network at T, and so the system undergoes a transition between two strong regimes (a strong-to-strong transition). The same behaviour is found in molecular~\cite{shi2018origin,shi2018common} and coarse grained models~\cite{russo2018water} of water, and is also a feature in two-state models of water~\cite{tanaka2020liquid}.

The low-$T$ scaling of the activation energy can be predicted within the mean-field theory of bonding. In this regime, the dynamics is controlled by defects, i.e. by the probability of having a broken bond, $1-p_b$. Starting from Eq.~\ref{eqn:pbond} we see that this quantity scales as

\begin{equation}
1-p_b\sim\exp\left(-\frac{\beta\epsilon}{2}\right)
\end{equation}

It is found that this prediction is followed only for systems with wide bond angles~\cite{smallenburg2013liquids}, where ultra-stable liquid behaviour is observed. For smaller patch widths, the geometrical constraints on bonding which are absent in the mean-field description, considerably increase the activation energy, even though the exact value depends on the details of the patch-patch interaction: for example, Ref.~\cite{roldan2017connectivity} finds $E_a\sim 1.35\epsilon$, and Ref.~\cite{de2006dynamics} finds $E_a\sim\epsilon$.  In brief, the activation energy can change from
$0.5 \epsilon$ for independent bond-breaking process, to $2\epsilon$ for the 
highly correlated simultaneous breaking of all four bonds.

The relation between the diffusion coefficient and the population of broken bonds in tetrahedral networks has been found very generally to scale as~\cite{de2006dynamics,roldan2017connectivity}

\begin{equation}\label{eqn:diffusion}
D\sim(1-p_b)^4
\end{equation}

In the mean-field theory of bonding $(1-p_b)^4$ is the population of particles with four broken bonds, i.e. monomers. It would then be natural to interpret Eq.~\ref{eqn:diffusion} as a relaxation process dominated by monomers, whose mobility is the highest because of the lack of energetic constraints. But it is found~\cite{roldan2017connectivity} that the number of monomers does not scale as $(1-p_b)^4$ because the mean-field approximation of bond-independence breaks down at low-$T$. This means that the probability of breaking a bond is not independent from the local energy. Instead it is observed that a particle with one broken bond is more likely to break another bond, and so on. Defects are thus not randomly distributed on the network, and entropy favors medium range spatial correlations between the defects. The spatial distribution of defects is linked to the distinction between network topologies upon which two-state models are derived~\cite{russo2014understanding}, and gives rise to so-called \emph{dynamic heterogeneities}, characterized by a finite dynamic correlation length and a non-gaussian statistics of displacements~\cite{shi2018common}. Dynamic heterogeneities are also observed in molecular models of water~\cite{la2004static}, and are a common occurrence in glasses where they originate from packing effects instead of bonding.

Despite bond-independence not holding at low-$T$, it is found that Eq.~\ref{eqn:diffusion} remains valid. Eq.~\ref{eqn:diffusion} tells that all bonded-networks are \emph{strong} glass formers, a fact well known in silica and water. It also explains dynamic anomalies. A non-monotonous dependence of $p_b$ in fact will cause a non-monotonous change in $D$. As we saw, optimal networks are characterized by an optimal density where the energy $u(\rho)$ has a minimum. This is where most bonds in the network are formed and will correspond to a diffusion minima; decreasing or increasing the density from its optimal value will then cause an acceleration of the dynamics.

It is interesting to note that bond breaking is not the only possible relaxation mechanism. In ultra-stable liquid systems, where the bond angle is very wide and where the condition of one-bond-per-patch is enforced, a more effective relaxation mechanism is given by bond switching, which causes the diffusion coefficient to scale as $D\sim(1-p_b)$.

\section{Conclusions}

Patchy particles are the ideal model to capture the general behaviour of 
tetrahedral networks, thanks to simple and physically motivated parameters that encode the geometric and energetic aspects of bonding.
Here we have attempted to elucidate some of the strange properties of water from the point of view of tetrahedral patchy-particle networks. 

In particular, with simple theories of bonding we have highlighted some key design principles of patchy particle models that help us to understand the behaviour of water.  
We started with the observation that water is indeed an empty liquid, i.e. a system with a liquid-gas coexistence confined to (relatively) low densities. This opens up the space in phase diagram where the anomalous properties of water can emerge. We have shown that a general mechanism for the existence of a second region of instability at high density is the optimal network condition, i.e. when geometrical constraints in the bonds do not allow a disordered network system to be arbitrarily compressed without sacrificing some of its bonding energy. Tuning this condition allows the instability to move relatively to the liquid-gas phase transition, giving rise to the popular water scenarios, including the two liquid critical points scenario, the stability-limit scenario, and the singularity-free and critical-point free scenarios.
We then considered water's thermodynamic anomalies, and saw how they also emerge naturally from the properties of the underlying tetrahedral network. The observation of water's anomalous behaviour is limited at high density by the nucleation of open crystalline structures. We have thus reviewed how the solid-liquid phase transition is affected by the parameters of patchy particle potentials, and shown that the bond angular width ($\theta_\text{max}$) is the key factor controlling the crystal-forming ability of these systems. We have also highlighted the conditions that disfavor the crystalline state, and illustrated two important properties that emerge from this suppression, i.e. ultra-stable liquids and crystal-clear liquid-liquid phase transitions.
We concluded with a discussion of the dynamical behaviour of patchy particle models, focusing on their glass-forming properties such as the transition from two strong regimes and the appearance of dynamic heterogeneities. All dynamical properties can be explained in terms of the underlying network, whose dynamics is controlled by bond-breaking events.

In conclusion, \emph{empty liquids} represent a broad category of fluids that derive their general properties from the existence of an underlying network of bonds, and that has found in water one of its prototypical and most inspiring members. Our knowledge of the number of systems that behave like empty liquids is constantly increasing and the study of their general properties is the subject of intense research.

\vspace{1cm}

\noindent
{\bf Acknowledgements}
\noindent
JR and FL acknowledge support from the European Research Council Grant DLV-759187.  FS acknowledges support from  MIUR-PRIN  (Grant No. 2017Z55KCW). FM acknowledges support from the STFC Hartree Centre's Innovation Return on Research programme, funded by the Department for Business, Energy and Industrial Strategy.

\vspace{5mm}

\end{document}